\documentclass[apj]{emulateapj}

\usepackage{ulem}
\usepackage{graphicx}
\usepackage{amsmath}
\usepackage{epsfig}
\usepackage{enumitem}
\usepackage{multirow}
\usepackage{courier}
\usepackage{longtable}
\usepackage{tabu}
\usepackage{booktabs}

\newcommand{\msol}{M{$_{\odot}$}}
\newcommand{\lsol}{L{$_{\odot}$}}

\begin{document}

\title{A Systematic Study of Mid-Infrared Emission from Core-Collapse Supernovae with SPIRITS }

\author{Samaporn Tinyanont\altaffilmark{1}}
\email{st@astro.caltech.edu}
\author{Mansi M. Kasliwal\altaffilmark{1}}
\author{Ori D. Fox\altaffilmark{2}}
\author{Ryan Lau\altaffilmark{1,3}}
\author{Nathan Smith\altaffilmark{4}}
\author{Robert Williams\altaffilmark{5}}
\author{Jacob Jencson\altaffilmark{1}}
\author{Daniel Perley\altaffilmark{6}}
\author{Devin Dykhoff\altaffilmark{7}}
\author{Robert Gehrz\altaffilmark{7}}
\author{Joel Johansson\altaffilmark{8}}
\author{Schuyler D. Van Dyk\altaffilmark{9}}
\author{Frank Masci\altaffilmark{9}}
\author{Ann Marie Cody\altaffilmark{10}}
\author{Tom Prince\altaffilmark{1}}
\altaffiltext{1}{California Institute of Technology, Pasadena, CA 91125, USA}
\altaffiltext{2}{Space Telescope Science Institute, 3700 San Martin Drive, Baltimore, MD 21218, USA}
\altaffiltext{3}{Jet Propulsion Laboratory, California Institute of Technology, 4800 Oak Grove Dr., Pasadena, CA 91109, USA}
\altaffiltext{4}{Steward Observatory, University of Arizona, Tucson, AZ 85721, USA}
\altaffiltext{5}{Space Telescope Science Institute, 3700 San Martin Drive, Baltimore, MD 21218, USA}
\altaffiltext{6}{Dark Cosmology Centre, Niels Bohr Institute, University of Copenhagen, Juliane Maries Vej 30, 2100 K{\o}benhavn {\O}, Denmark}
\altaffiltext{7}{Minnesota Institute for Astrophysics, School of Physics and Astronomy, University of Minnesota, 116 Church Street, S. E., Minneapolis, MN 55455, USA}
\altaffiltext{8}{Benoziyo Center for Astrophysics, Weizmann Institute of Science, 76100 Rehovot, Israel}
\altaffiltext{9}{Infrared Processing and Analysis Center, California Institute of Technology, M/S 100-22, Pasadena, CA 91125, USA}
\altaffiltext{10}{NASA Ames Research Center, Moffett Field, CA 94035, USA}

\begin{abstract}

We present a systematic study of mid-infrared (mid-IR) emission from 141 nearby supernovae (SNe) observed with the InfraRed Array Camera (IRAC) on the \textit{Spitzer} Space Telescope. These SNe reside in one of the 190 galaxies within 20\,Mpc drawn from the ongoing three-year SPIRITS program. Both new SPIRITS observations and data from other programs available in the archive are used in this study. We detect 8 Type Ia SNe and 36 core-collapse SNe. All Type I SNe fade and become undetectable within 3 years of explosion. About 22$\pm$11\% of Type II SNe continue to be detected at late-times with five events detected even two decades after discovery. Dust luminosity, temperature, and mass are obtained by fitting the spectral energy distributions using photometry with IRAC bands 1 and 2. The dust mass estimate is a lower limit as the dust cloud could be optically thick or there could be cooler dust hiding at longer wavelengths. The estimate also does not distinguish between pre-existing and newly produced dust. We observe warm dust masses between $10^{-2}$ and $10^{-6}$ \msol\, and dust temperatures from 200\,K to 1280\,K. 
We present detailed case studies of two extreme Type II-P SNe: SN\,2011ja and SN\,2014bi.  SN\,2011ja was over-luminous ([4.5] = $-$15.6\,mag) at 900\,days post-explosion accompanied by a monotonic growth of the dust mass. This suggests either an episode of dust formation similar to SN\,2004et and SN\,2004dj, or an intensifying CSM interactions heating up pre-existing dust. SN\,2014bi showed a factor of 10 decrease in dust mass over one month suggesting either an episode of dust destruction or a fading source of dust heating. A re-brightening in the mid-IR light curve of the Type Ib SN\,2014C coinciding with a rise in the dust mass indicates either an episode of dust production perhaps via CSM interactions or more pre-existing dust getting heated up by the CSM interactions. This observation adds to a small number of stripped-envelope SNe that have mid-IR excess as has been previously reported in the case of SN\,2006jc. The observed dust mass and the location of the CSM interactions suggest that the CSM shell around SN\,2014C is originated from an LBV-like eruption roughly 100 years before the explosion. We also report detections of SN\,1974E, SN\,1979C, SN\,1980K, SN\,1986J, and SN\,1993J more than 20 years post-explosion. The number of outlying SNe identified in this work demonstrates the power of late time mid-IR observations of a large sample of SNe to identify events with unusual evolution. 
	 
\end{abstract}

\section{Introduction}
The mid-IR evolution of core-collapse supernovae (CCSNe) remains relatively uncharacterized, except in a few specific events. A more comprehensive study of a large sample of CCSNe across all types is essential to delineate the temporal evolution of a massive star after its explosive death. One of the advantages of the mid-IR is that it can trace emission from warm dust in the ejecta of a CCSN or in the circumstellar material (CSM) that is related to the progenitor star.
CCSNe have long been proposed as major sources of dust production, especially in high-redshift galaxies where other possible dust factories (e.g. stellar winds of AGB stars) could not be in operation either because those galaxies were too young to form AGB stars or the stellar metallicity in that epoch was inadequate to form dust particles (\citealp{gall2011, cherchneff2014} and references therein). Theoretical models predict the mass of dust produced in CCSNe to be around 0.1 to 1 \msol\,, which is adequate for CCSNe to contribute significantly to the observed dust content in the early universe \citep{nozawa2003, nozawa2008}. Dust characteristics of SNe, such as grain size, mass, and temperature can be quantified by analyzing the Spectral Energy Distribution (SED) and the light curve of an SN (see, for example, \citealp{fox2010, fox2011, fox2013,szalai2013}). \cite{gehrz1990} described three distinct signatures of dust formation that were observed in SN\,1987A and that are expected to be generally observable in other SNe that form dust. There are: (1) Optical spectral lines are asymmetrically blueshifted because the newly formed dust absorbs light coming from the receding side of the ejecta. (2) The decline rate of the optical light curve increases because the new dust causes more extinction in those wavebands. (3) The emission in the mid-IR becomes relatively brighter than that in optical and near-IR due to emission from warm dust. 

A number of SNe have been identified as dust producers through observations in the mid-IR. For example, \cite{helou2013} and \cite{ergon2014} conducted mid-IR studies of the Type IIb SN\,2011dh in Messier\,51 (see \citealp{filippenko1997} for more information on supernova classification.) The time evolution of its SED showed that the mid-IR emission became relatively brighter than optical emission at late times. The light curve also showed excess mid-IR emission that could not be explained by an IR echo alone, indicating an additional dust heating mechanism or the formation of new dust.  \cite{meikle2011} and \cite{szalai2011} analyzed the Type II-P SN\,2004dj and found a significant re-brightening $\sim$ 400 days post-explosion. Similarly, \cite{fabbri2011} studied the Type II-P SN\,2004et and observed re-brightening at 1000 days after the explosion. In both cases, they infer new dust formation by observing red-wing attenuation in optical spectral lines and the sharp decline in optical light curves. 
A mid-IR population study of 12 Type II-P SNe by \cite{szalai2013} found warm dust emission with an inferred dust mass of $\sim 10^{-3}$\msol\,.
\cite{fox2011,fox2013} conducted a survey on all known 68 Type IIn events. These are CCSNe with narrow emission lines that are indicators of circumstellar interactions that either produce new dust or heat up pre-existing dust. They detected late-time emission from 10 targets with an inferred dust mass of up to $10^{-2}$ \msol\,. 
In addition to Type IIn SNe whose CSM interactions were discovered early on, recent late-time observations reveal that many SNe interact with the CSM later in their evolution, providing additional channels for dust production. \cite{andrews2015} presented spectroscopic evidence for CSM interactions in the Type II-P SN\,2011ja at 64-84 days post explosion accompanied by an episode of dust production between the forward and the reverse shocks. These interactions with dust productions are also observed in a small number of Type Ibc SNe, such as SN\,2006jc (see e.g. \citealp{smith2008,mattila2008,sakon2009} and references therein.) 

The mass of newly formed dust observed in a number of CCSNe is significantly lower than dust masses predicted by theoretical models, falling between $10^{-3}$ to $10^{-4}$ \msol\, per event instead of 0.1-1 \msol\, (see in addition, \citealp{gall2014, andrews2010}). 
It is possible that there is a considerable mass of dust that is too cold to be detected in the mid-IR. Recent studies of SN\,1987A and other supernova remnants (SNRs) reveal a large amount of dust emitting in the far-IR and submillimeter. \cite{matsuura2015} used the \textit{Herschel Space Observatory} to observe SN\,1987A and deduced, depending on dust composition, 0.5-0.8 \msol\, of cold dust in the remnant. \cite{indebetouw2014} used the Atacama Large Millimeter/Submillimeter Array (ALMA) to observe SN\,1987A at high spatial resolution and confirmed that $>$ 0.2 \msol\, of dust was formed in the inner ejecta and was neither from interactions with CSM as suggested by some authors (e.g. \citealp{bouchet2014}) nor pre-existing. This dust mass is orders of magnitude higher than the $\sim 10^{-4}$ \msol\, reported by \cite{wooden1993} two years post-explosion. This suggests a possibility that a large bulk of dust is condensed later in the evolution of an SN during the remnant phase. Alternatively, \cite{dwek2015} suggested that the dust was hidden in the optically thick part of the ejecta at early times ($<$ 1000d), and not observed until later. The presence of about 0.1 \msol\, of cold dust was also suggested by far-infrared and submillimeter observations of the nearby SNRs, Cassiopeia A \citep{barlow2010} and the Crab Nebula \citep{gomez2010}. Recently, \cite{lau2015} have reported that 0.02 \msol\, of ejecta-formed dust has survived in the 10,000 yr-old SNR Sgr A East. These results demonstrate that a large fraction of the dust formed in the inner core of a SN can survive the reverse shocks and later be dispersed into the ISM. Although a large fraction of newly formed dust appears to emit at longer wavelengths, the mid-IR band is still a powerful tool for identifying dust formation events. 

This work seeks to study a comprehensive sample of different types of CCSNe observed by \textit{Spitzer} at a variety of epochs after explosion in order to construct an overview of the mid-IR time evolution of these events. This work complements the mid-IR light curve templates that have already been compiled for Type Ia SNe \citep{johansson2014}. The collage of light curves from a number of well-observed SNe Ia shows that these events are homogeneous in their time evolution. In this paper, we compile a light curve collage for core-collapsed supernovae in order to identify the typical time evolution of these events and to potentially uncover some unusual SNe. 
 A simple one component, graphite dust model is fitted to the SED to extract the dust temperature and mass. Many SNe are observed at several epochs, allowing us to trace the time evolution of dust parameters as well. 
 In section \ref{sec:obs}, we discuss SPIRITS in more detail along with the SNe sample in galaxies covered by SPIRITS. We present the demographics of the SNe and light curve collages for Type Ib/c and II SNe and identify and discuss briefly events which are outliers in section \ref{sec:results}. Dust parameters for each observation are extracted and presented in this section. 
In section \ref{sec:cases} we present four intriguing case studies: the over-luminous Type II SN\,2011ja, the re-brightening Type Ib SN\,2014C, the under-luminous and extremely red Type II-P SN\,2014bi, and senior SNe with mid-IR emission detected more than 20 years after explosions. 


\section{Observations}\label{sec:obs}
\subsection{Supernovae Sample and \textit{Spitzer}/IRAC Photometry}

 The SPitzer InfraRed Intensive Transient Survey (SPIRITS; Kasliwal et al. 2015, in prep.) targets 190 nearby galaxies within 20 Mpc to a depth of 20 mags on the Vega system. The observations are performed in the 3.6 $\mu$m and 4.5 $\mu$m bands of the InfraRed Array Camera (IRAC, \citealp{fazio2004}) on board the \textit{Spitzer} Space Telescope \citep{werner2004, gehrz2007}. Magnitudes in both bands are denoted [3.6] and [4.5] hereafter. All magnitudes are on the Vega system and the conversion $m_{\rm AB} - m_{\rm Vega}$ is +2.78 for [3.6] and +3.26 for [4.5] computed from the zeropoints given in the IRAC Instrument Handbook. The SPIRITS galaxy sample has hosted 141 SNe in total since 1901 and these SNe have been observed by \textit{Spitzer} in at least one epoch.\footnote{The SNe list is obtained from the IAU Central Bureau for Astronomical Telegram at \url{http://www.cbat.eps.harvard.edu/lists/Supernovae.html}.} With the combination of archival data and new SPIRITS observations, 44 SNe are detected: 8 are Type Ia, 9 are Type Ib/c, and the remaining 27 are Type II. Seven new SNe exploded in this galaxy sample during 2014. All of them are observed and 6 are detected. The exception, SN\,2014bc, is inside its host's nucleus and all observations of it are saturated. Two events, SN 2014J and SN 2014dt, are Type Ia which are outside the scope of this paper. Refer to \cite{johansson2014} and \cite{fox2015} respectively for detailed discussions about these events.
 

 We used a combination of data from new SPIRITS observations and from other programs, both publicly available from the \textit{Spitzer} Heritage Archive, to study mid-IR emission from the SNe of Type Ib/c and different subtypes of Type II. We used Post Basic Calibrated Data (\texttt{pbcd}) which had been coadded and calibrated by the standard pipeline. The SPIRITS pipeline performs PSF matching image subtraction using stacks of archival \textit{Spitzer} data, generally ranging from 2004-2008, as references. The actual range of epochs used as reference depends on the availability of the data for each host galaxy. 
 
 For 34 SNe that exploded after this range of epochs, there is no SN light present in the reference frames. In this case the aperture photometry is performed on the subtracted image to eliminate the flux contribution from the galaxy background, which is bright and spatially varying. This methodology is applied to every epoch of each SN for which there exists a pre-explosion reference image. We checked our photometry against IRAC photometry of SN\,2011dh presented by \cite{ergon2014} between 18 to 1061 days post-explosion. The results agree to within 5\%. In the case of SN\,2011ja, the SN location is near the edge of the frame and the pipeline fails to produce good subtraction images. We instead use aperture photometry on all frames, then subtract the average baseline flux in pre-explosion images from each epoch in which the SN appears. This method provides good agreement with the image subtraction method in cases where we have high quality subtractions. 12 SNe of Type Ib/c and II that are detected and have reliable photometry (i.e. not too close to the host's nucleus) are bold in Table \ref{sn_data}. 
 
 For SNe whose reference frames were contaminated with SN light, we searched for a distinct point source at the location of a SN in the science frame to judge whether or not we had a detection. In cases judged to be detections, aperture photometry with sky background subtraction was performed on the science frames. The aperture corrections from the IRAC Instrument Handbook were then applied. Because the galaxy background could not be completely eliminated in these cases, the photometric data had larger uncertainties especially at late times when the SNe light was comparable in brightness with the background. 
 In order to ascertain the detection, the thumbnails of every observation of every SN were visually vetted. We excluded SN impostors SN\,1997bs and SN\,2008S \citep{kochanek2012}. We also excluded SNe that were located near their host's nucleus so that all observations of them were saturated (e.g. SN\,2004am, 2008iz, and 2014bc). By doing image subtraction, we were able to identify mid-IR light from old and slowly evolving SNe that escaped notice of previous works because they were very dim in comparison to the background (see e.g. Table 1 in \citealp{fox2011}). All SNe with at least one detection with \textit{Spitzer}, either with the new data from SPIRITS or from other programs, are presented in Table \ref{sn_data}. 
 
\begin{deluxetable*}{llllllll}
\caption{All SNe in this sample detected with \textit{Spitzer} for at least one epoch. Bold are CCSNe with pre-explosion images for image subtraction.}
\label{sn_data} \\ \toprule
Name & Type & Discovery & RA (J2000) & Dec (J2000) & Host & d (Mpc) & Reference \\ \midrule
1974E &		 II &		 1974-03-21 &		 180.475 &		 -18.866667 &		 NGC 4038 			&	 20.04   &	---	  \\
1979C &		 II-L &		 1979-04-19 &		 185.744292 &		 15.797694 &		 NGC 4321 		&	 17.95   & \cite{dwek1983} \\
1980K &      II-L &     1980-10-28  &      308.875292  &          60.106611  &    NGC 6946       & 6.67 & \cite{dwek1983,dwek1983b} \\
1986J &		 II &		 1986-08-21 &		 35.630417 &		 42.3325 &		 NGC 891 			&	  8.36  &		\cite{cappellaro1986, milisavljevic2008} \\
1993J &		 IIb &		 1993-03-28 &		 148.854167 &		 69.020278 &		 NGC 3031 		&	  3.63   &		\cite{filippenko1993}  \\
1999bw &		 IIn &		 1999-04-20 &		 154.945042 &		 45.526389 &		 NGC 3198 &		 11.64   & \cite{sugerman2004}	  \\
2003gd &		 II-P &		 2003-06-12 &		 24.177708 &		 15.739139 &		 NGC 628 &		  8.59   &		 \cite{sugerman2006, meikle2007} \\
2003hn &		 II-P &		 2003-08-25 &		 56.150417 &		 -44.630278 &		 NGC 1448 &		 13.43   &	\cite{salvo2003, szalai2013}	  \\
2003J &		 	II-P &		 2003-01-11 &		 182.7405 &		 50.4755 &		 NGC 4157 			&	 14.32   &	\cite{ayani2003, szalai2013}	  \\
2004cc &		 Ic &		 2004-06-10 &		 189.143333 &		 11.242444 &		 NGC 4568 &		 33.73   &		\cite{foley2004}  \\
2004dj &		 II-P &		 2004-07-31 &		 114.320917 &		 65.599389 &		 NGC 2403 &		  3.18   &	\cite{meikle2011,szalai2011}	 \\
2004et &		 II &		 2004-09-27 &		 308.855542 &		 60.121583 &		 NGC 6946 &		  6.67   & \cite{kotak2009, fabbri2011}		  \\
2004gn &		 Ib/c &		 2004-12-01 &		 188.550417 &		 2.659556 &		 NGC 4527 	&		  9.68  &	\cite{pugh2004}	  \\
2005ae &		 IIb? &		 2005-02-01 &		 119.566 &		 -49.855528 &		 ESO 209-9 &		 12.30   &		\cite{filippenko2005}  \\
2005af &		 II-P &		 2005-02-08 &		 196.183583 &		 -49.566611 &		 NGC 4945 &		  3.80   &		\cite{kotak2006, szalai2013}  \\
2005at &		 Ic &		 2005-03-15 &		 287.473208 &		 -63.823 &		 NGC 6744 			 & 7.69  &		 \cite{kankare2014} \\
2005cs &		 II-P &		 2005-06-28 &		 202.469917 &		 47.176583 &		 NGC 5194 &		  7.66   &		\cite{szalai2013}  \\
2005df &		 Ia &		 2005-08-04 &		 64.407708 &		 -62.769306 &		 NGC 1559 &		 10.76   &		\cite{salvo2005, diamond2015}  \\
2006mq &		 Ia &		 2006-10-22 &		 121.551625 &		 -27.562611 &		 ESO 494-G26 &	 11.12   &	\cite{duszanowicz2006}	  \\
2006ov &		 II-P &		 2006-11-24 &		 185.480417 &		 4.487972 &		 NGC 4303 	&		 18.45   &	\cite{szalai2013}	  \\
2006X &		 	Ia &		 2006-02-04 &		 185.724958 &		 15.809194 &		 NGC 4321 	&	 17.95   &	\cite{lauroesch2006, wang2008}	  \\
2007it &		 II &		 2007-09-13 &		 214.606792 &		 -43.381611 &		 NGC 5530 &		 12.94   &	\cite{andrews2011}	  \\
2007sr &		 Ia &		 2007-12-18 &		 180.47 &		 -18.972694 &		 NGC 4038 	&		 20.04   &	\cite{pojmanski2008}	  \\
2009hd &		 II-L &		 2009-07-02 &		 170.070667 &		 12.979611 &		 NGC 3627 &		 10.81   &	\cite{elias2011}	  \\
2010br &		 Ib/c &		 2010-04-10 &		 180.795625 &		 44.528639 &		 NGC 4051 &		 22.39   &	\cite{maxwell2010}	  \\
\textbf{2011dh} &		 IIb &		 2011-06-01 &		 202.521333 &		 47.169667 &		 NGC 5194 &		  7.66   &	 \cite{helou2013, ergon2014}  \\
2011fe &		 Ia &		 2011-08-24 &		 210.774208 &		 54.273722 &		 Messier 101 		&		  7.38   &	 \cite{mcclelland2013}	  \\
\textbf{2011ja} &		 II-P &		 2011-12-10 &		 196.296333 &		 -49.524167 &		 NGC 4945 &		  3.80   &		\cite{andrews2015}  \\
2012aw &		 II-P &		 2012-03-16 &		 160.974 &		 11.671639 &		 NGC 3351 		&	  8.79  &		\cite{siviero2012}  \\
2012cc &		 II &		 2012-04-29 &		 186.736708 &		 15.045972 &		 NGC 4419 &		 13.49   &	\cite{marion2012}	  \\
2012cg &		 Ia &		 2012-05-17 &		 186.803458 &		 9.420333 &		 NGC 4424 	&		  7.31   &		 \cite{marion2012b} \\
2012fh &		 Ib/c &		 2012-10-18 &		 160.891875 &		 24.891389 &		 NGC 3344 &		 19.05   &	\cite{nakano2012}	  \\
\textbf{2013ai} &		 II &		 2013-03-01 &		 94.076458 &		 -21.375806 &		 NGC 2207 &		 14.26   &	\cite{conseil2013}	  \\
\textbf{2013am} &		 II &		 2013-03-21 &		 169.737292 &		 13.063722 &		 NGC 3623 &		 18.54   & \cite{nakano2013}		  \\
\textbf{2013bu} &		 II &		 2013-04-21 &		 339.259042 &		 34.401444 &		 NGC 7331 &		 13.12   &	\cite{itagaki2013}	  \\
\textbf{2013df} &		 IIb &		 2013-06-07 &		 186.622208 &		 31.227306 &		 NGC 4414 &		 20.51   &  \cite{morales2014}	; Szalai (In Prep.)	  \\
\textbf{2013dk} &		 Ic &		 2013-06-22 &		 180.469667 &		 -18.87175 &		 NGC 4038 &		 20.04   &	\cite{elias2013}	 \\
2013ee &		 II &		 2013-07-13 &		 150.486792 &		 55.695556 &		 NGC 3079 &		 14.86   &	\cite{cortini2013}	  \\
\textbf{2013ej} &		 II-P &		 2013-07-25 &		 24.200667 &		 15.758611 &		 NGC 628 &		  8.59   &		\cite{bose2015}  \\
\textbf{2014bi} &		 II-P &		 2014-05-31 &		 181.512458 &		 47.492639 &		 NGC 4096 &		 11.27   &		\cite{kumar2014}  \\
\textbf{2014C} &		 	Ib &		 2014-01-05 &		 339.273333 &		 34.408861 &		 NGC 7331 	&	 13.12   &	\cite{kim2014, milisavljevic2015}	  \\
\textbf{2014df} &		 Ib &		 2014-06-03 &		 56.099958 &		 -44.668917 &		 NGC 1448 &		 13.43   &	\cite{monard2014}	  \\
2014dt &		 Ia-p &		 2014-10-29 &		 185.489875 &		 4.471806 &		 NGC 4303	&		 18.45   &	\cite{fox2015}	  \\
2014J &		 	Ia &		 2014-01-21 &		 148.925583 &		 69.673889 &		 NGC 3034	&     3.52   &	\cite{johansson2014}	 \\
\textbf{2014L} &		 	Ic &		 2014-01-26 &		 184.702833 &		 14.412083 &		 NGC 4254	&	 16.83   &	\cite{yamaoka2014}	  \\
\midrule
\end{deluxetable*}

\subsection{Near-Infrared Photometry}
As part of the follow-up campaign accompanying SPIRITS, some of the recent SNe listed in Table \ref{sn_data} were observed in near-IR from multiple observatories. 
SN\,2013bu, SN\,2013df, SN\,2014C, and SN\,2014bi were observed in the J, H, and Ks bands using the Two Micron All Sky Survey spectrometer (2MASS; \citealp{milligan1996}) on the 60-inch telescope at the Mount Lemmon Observational Facility. The science frame and the surrounding sky are observed alternately for sky subtraction. No SNe were detected, so the limiting magnitudes (3$\sigma$) are given in Table \ref{nir_sn}. 
SN\,2014C were observed in the J, H, and Ks bands using the 4' × 4' Nordic Optical Telescope near-infrared Camera and spectrograph (NOTCam; \citealp{abbott2000}) at the Nordic Optical Telescope under the programme 51-032 (PI: Johansson). We used the wide field imaging mode and beam-switching to guarantee successful sky subtraction.
Data reduction was performed with the NOTCam Quick-Look reduction package based on IRAF. PSF fitting photometry was performed on the sky-subtracted frames, calibrated using 2MASS stars in the field.
SN\,2014C was observed in the Ks band using the Wide Field InfraRed Camera (WIRC; \citealp{wilson2003}) on the 200-inch Hale Telescope at Palomar Observatory. The data were sky subtracted and calibrated using 2MASS stars in the field. 
The resulting photometry are listed in Table \ref{nir_sn}. 

\begin{deluxetable*}{llllll}
\caption{A table summarizing all near-infrared photometry of the SNe in Table \ref{sn_data}. Magnitudes listed are Vega, calibrated using 2MASS stars.}
\label{nir_sn} \\ \toprule 
MJD & Epoch  & $J$ & $H$ & $Ks$ & Instrument \\
& days & & & & \\ \midrule
\multicolumn{6}{c}{SN\,2013bu (II)} \\ \midrule
56809 & 405.5 & \textgreater 16.1        & \textgreater15.5       & \textgreater15.1       & MLOF/2MASS \\
56901 & 497.5 & \textgreater16.2        & \textgreater15.5       & \textgreater15.3       & MLOF/2MASS \\
56931 & 527.5 & \textgreater16.1        & \textgreater15.3       & \textgreater14.9       & MLOF/2MASS \\
56950 & 546.5 & ---          & ---         & \textgreater19.0       & P200/WIRC  \\
56962 & 558.5 & \textgreater16.1        & \textgreater15.1       & \textgreater14.5       & MLOF/2MASS \\
\midrule \multicolumn{6}{c}{SN\,2013df (IIb)} \\ \midrule
56717 & 266.5 & \textgreater15.9        & \textgreater14.9       & \textgreater14.9       & MLOF/2MASS \\
57023 & 572.5 & \textgreater14.5        & \textgreater13.7       & \textgreater13.9       & MLOF/2MASS \\
\midrule \multicolumn{6}{c}{SN\,2014C (Ib)} \\ \midrule
56809 & 146.5 & \textgreater13.4        & \textgreater12.7       & \textgreater12.1       & MLOF/2MASS \\
56901 & 238.5 & \textgreater13.9        & \textgreater13.0       & \textgreater12.6       & MLOF/2MASS \\
56931 & 268.5 & \textgreater13.5        & \textgreater12.8       & \textgreater12.3       & MLOF/2MASS \\
56950 & 287.5 & ---          & ---         & 14.38(0.06) & P200/WIRC \\
56962 & 299.5 & \textgreater13.5        & \textgreater12.5       & \textgreater12.3       & MLOF/2MASS \\
57292 & 630.5 & 18.0 (0.1) &             16.6 (0.1) &                  14.9 (0.1)                   & NOT/NOTCam \\
\midrule \multicolumn{6}{c}{SN\,2014bi (II-P)} \\ \midrule
57113 & 304.5 & ---          & \textgreater14.8       & \textgreater15.3       & MLOF/2MASS \\
\midrule
\end{deluxetable*}


\section{Results}\label{sec:results}
\subsection{Demographics} 

\begin{figure}
	\includegraphics[width = \linewidth]{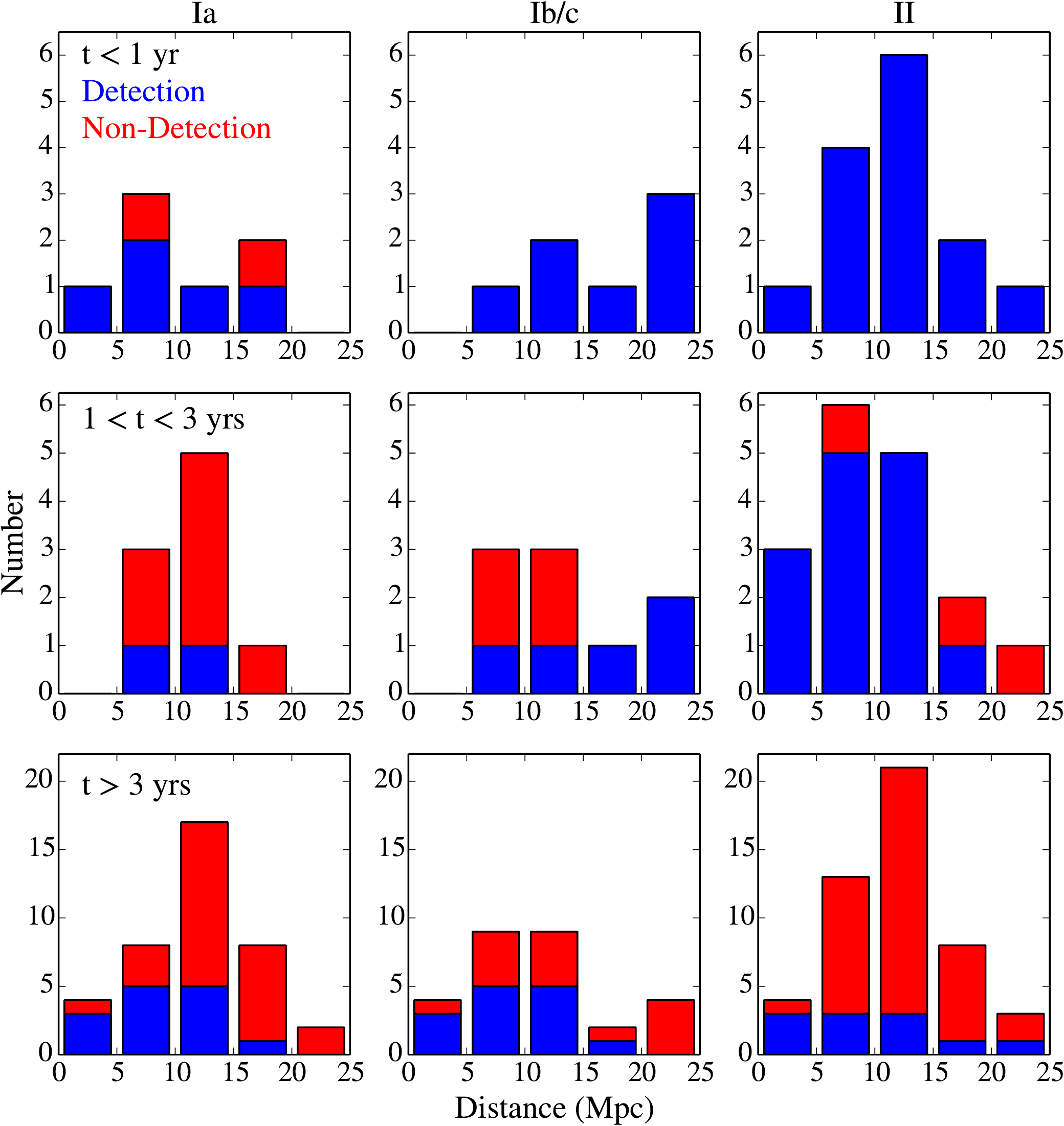}
	\caption{Stacked histograms showing detection statistics in our SN sample. Each column represents one type of SN and each row represents one bin of time after first discovery. Due to the small number of counts in our sample, we follow the procedure described by \cite{cameron2011} based on a Bayesian approach to estimating uncertainties. }
	\label{det_hist}
\end{figure}

The detection statistics for each type of SNe are presented in Fig. \ref{det_hist}. We consider separately three time bins after discovery: less than one year, one to three years, and more than three years. In each bin, if a SN is observed with at least one detection, it is marked as detected even though it might fade to non-detection later in the same bin. Otherwise, if all observations of a SN in that bin result in non-detections, then that SN is marked as a non-detection. The number of detections and non-detections of each type of SNe in each bin is then tallied up. The error estimates for small number statistics are done using the method described \cite{cameron2011} based on Bayesian statistics.

The distance to a SN's host galaxy may affect whether we detect the event. When one considers a given type of SNe, events further away tend to fall off of the detection limit first. However, across different types, the distance distributions in our sample are similar. Therefore, the different detection statistics we find for each type of SNe are not only on account of the Malmquist bias (the preferential detection of intrinsically bright objects), i.e. the relative fraction of detected Type II SNe versus Type Ia at late time is not due to the distance distribution. The results reflect the intrinsic difference in mid-IR evolution between CCSNe and Type Ia SNe.

The results show that almost all SNe across all types are detected within 1 year after discovery. The only two exceptions are Type Ia SN\,2004W and SN\,2011B. Both SNe were undetected in both IRAC channels, six and ten months respectively after their discovery dates. Overall, the mid-IR emission from most Type Ia SNe in our sample is shorter lived compared to CCSNe. We note that our sample does not include Type Ia events with CSM interactions (Ia-CSM), in which case mid-IR emissions can live longer (see e.g. SN\,2005gj and SN\,2002ic, \citealp{fox2013b}; SN 2014dt \citealp{fox2015}).  

Between 1 and 3 years after discovery, the fractions of Type Ia and Ib/c detected dropped dramatically, to 22$\pm$24\% and 55$\pm$28\% respectively. However, most of Type II SNe (82$\pm$18\%) continued to be detected. As discussed above, this is not a result of the distance bias in the sample. In fact, 3 out of 4 detected Type Ib/c events in this bin are further away than all 4 non-detected ones.

Finally, after 3 years, none of the Type Ia and Ib/c were detected. However, 22$\pm$11\% of the Type II events remained detected, mostly in the 4.5 $\mu$m band. These long-lived events are SN\,1974E (II), SN\,1979C (II-L), SN\,1986J (IIn), SN\,1993J (IIb), SN\,1999bw (IIn; \citealp{filippenko1999, kochanek2012}), SN\,2004dj (II-P), SN\,2004et (II-P), 2007it (II-P), 2009hd (II-L), and 2011ja (II-P; \citealp{andrews2015}). 
Additional discussion of a well-sampled subset of these old SNe follows in section \ref{old}. 
It is worth noting that the SN impostors, SN\,1997bs and SN\,2008S, were detected more than 3 years post-discovery, outlasting many other real Type II events even though they are not particularly nearby. 
Because late-time mid-IR emission comes mostly from warm dust, this data set allows us to probe evolution of dust properties in these CCSNe. 


\subsection{Light curves} 
\subsubsection{Type Ib/c Supernovae} 
\begin{figure}
	\includegraphics[width = \linewidth]{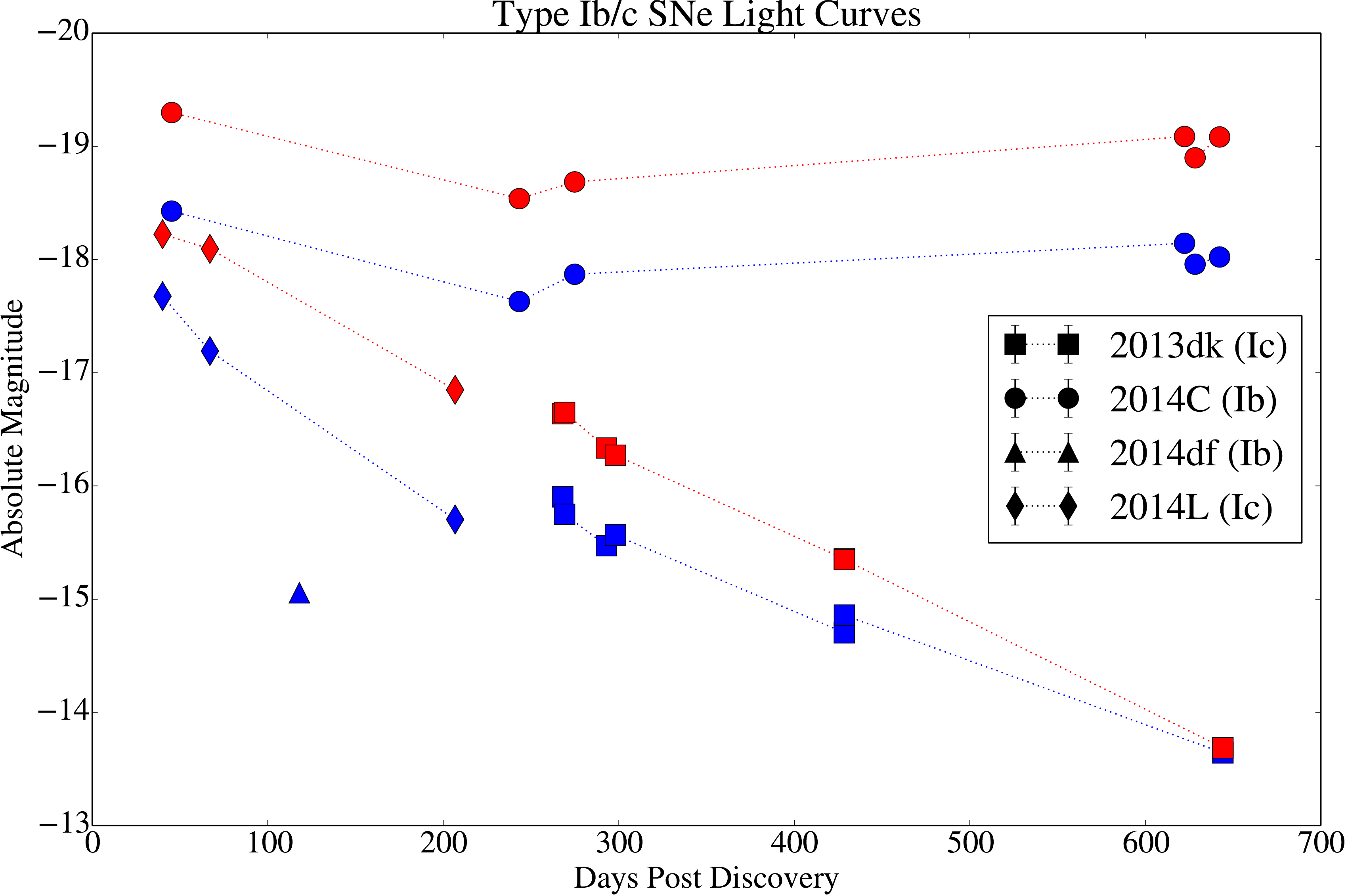}
	\caption{A collage of light curves of Type Ib/c SNe with photometry based on image subtraction. Red and blue markers represent [4.5] and [3.6] respectively. The error bars are smaller than the plotting symbols.}
	\label{fig:Ibc_collage}
\end{figure}

The light curves of all four Type Ib/c SNe for which we have image-subtracted photometry are shown in Figure~\ref{fig:Ibc_collage}. The data coverage spans epochs from 40 to 700 days after discovery, with multi-epoch data for SN\,2013dk (Ic), SN\,2014L (Ic), and SN\,2014C (Ib). The projected peak brightnesses range from [4.5] = -18.3 mags and [3.6] = -18.3 mags for SN\,2014L to [4.5] = -19.7 and [3.6] = -19 for SN\,2014C. The decline rate for SN\,2014C is 0.5 mags/100 days in both channels within 200 days after the explosion. The two Type Ic events decay at roughly the same rates in [4.5]: about 0.8 mags/100d within 450d. However, the [3.6] decline rates for these events are quite different. SN\,2014L decays for 1 mag/100d out to 200 days, while SN\,2013dk only decays at the rate of 0.68/100d between 250 to 430d. SN\,2014df (Ib) is about 3.5 mags dimmer than SN\,2014C, another Ib event, at the same epoch, but the lack of concurrent photometry yields no color information. No further observations of this galaxy have been made in 2015. The most interesting Type Ib/c SN observed in our sample is SN\,2014C, which re-brightens by 0.15 mags in 4.5 $\mu$m and 0.24 mags in 3.6 $\mu$m around 250d. It was observed again more than a year later at 620d to be even brighter in both bands. This peculiar time-evolution in the mid-IR sets this event apart from other SNe in this sample. Further discussion of this SN along with NIR spectra from Keck/MOSFIRE are presented in Sec. \ref{2014C}.
 We also note that, apart from a peculiar Type Ib SN\,2006jc which has CSM interactions (see \citealp{smith2008}, \citealp{mattila2008}), dust emission has not been detected for any other Type Ib/c events. 
\subsubsection{Type II Supernovae}
\begin{figure}
	\includegraphics[width = \linewidth]{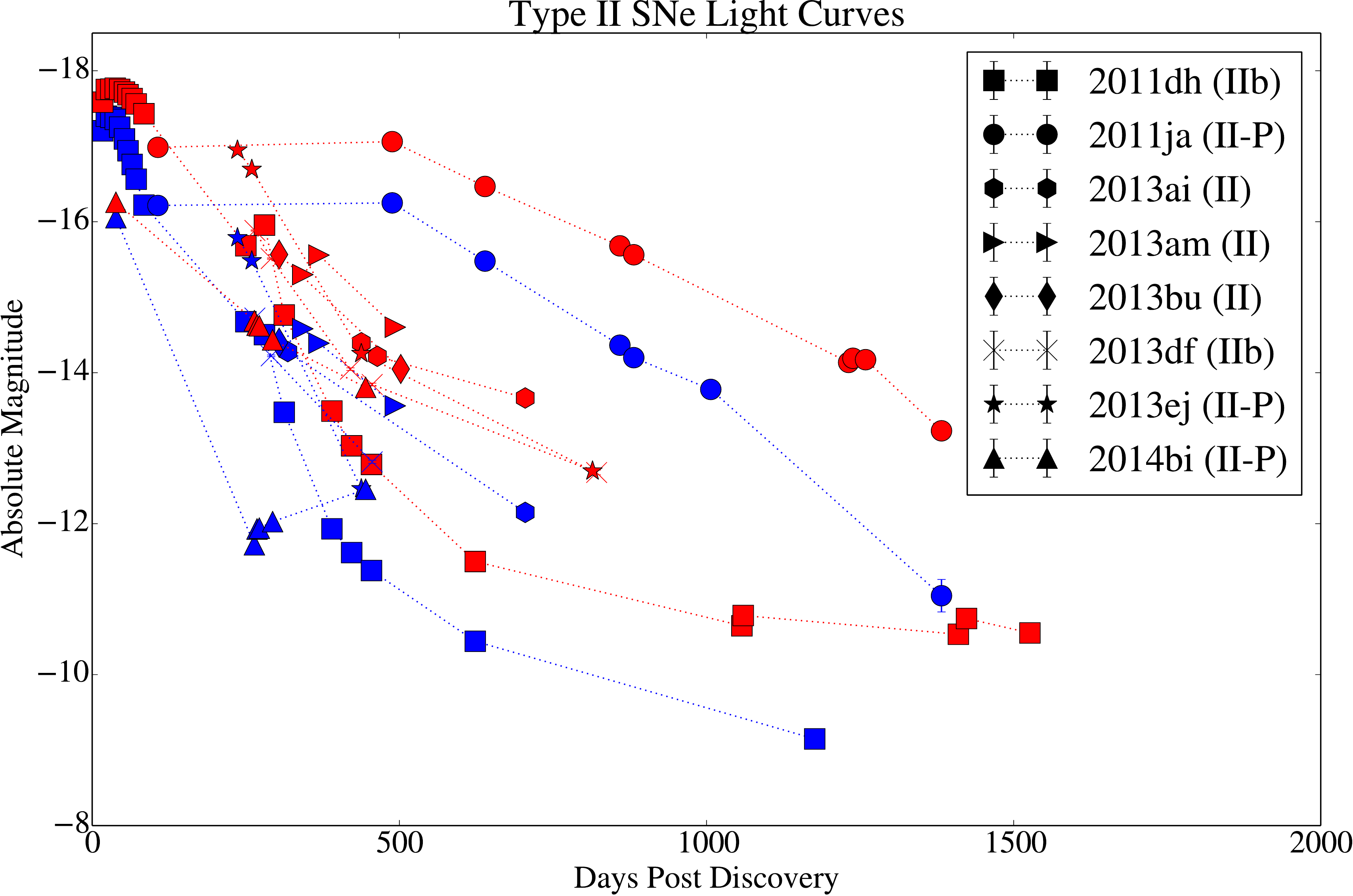}
	\caption{A collage of light curves of Type II SNe with photometry based on image subtraction. Red and blue markers represent [4.5] and [3.6] respectively. The error bars are smaller than the plotting symbols. }
	\label{fig:II_collage}
\end{figure}

The light curves of all 8 Type II SNe for which we have image-subtraction photometry are presented in Figure~\ref{fig:II_collage}. SN\,2011dh has the most well sampled light curves covering epochs from pre-maximum to 1200 days after the detection, providing a template with which other SNe can be compared. \cite{helou2013} interpolated [4.5] data to fill in the visibility gap between 90 and 250 days and found that 2011dh left its initial plateau into a decay at 150$\pm$25 days post-explosion. Between 200 to 400 days, SN\,2011dh has steep decline rates of 1.96 and 1.57 mags/100d in 3.6 $\mu$m and 4.5 $\mu$m. After 400 days, the light curves in both bands plateau out again to the decline rate of 0.57 and 0.75 mags/100d in 3.6 $\mu$m and 4.5 $\mu$m. After 600 days, the decline rates decrease to only 0.02 mags/100d in both channels. In these epochs after 100 days, \cite{helou2013} reported that an IR echo alone cannot account for the flux and that dust formation, additional heating mechanisms, a nearby interstellar cloud, or line or band emission must contribute. \cite{ergon2014} reported the fractional increase of 3.6 $\mu$m, 4.5 $\mu$m, and K bands luminosity compared to optical in conjunction with an increase in the decline rate of the optical pseudo-bolometric light curve. They also observed small blueshifts in the [OI] and [MgI] lines at 415 days. These results suggest that new dust is being created inside the expanding shell of ejecta attenuating optical light coming from the receding side of the shell. 

The collage in Fig. \ref{fig:II_collage} shows heterogeneity in the Type II light curves in comparison with 2011dh. Some SNe---such as SN\,2013bu (II), SN\,2013df (IIb), and SN\,2013ej (II-P)---have similar temporal evolution as that of SN\,2011dh in terms of the absolute magnitudes, which are within 1 mag from 2011dh, and the decline rates. SN\,2013ai shows similar plateauing after 400 days in 4.5 $\mu$m to the decline rate of 0.23 mags/100d, but there is no sufficient coverage in the 3.6 $\mu$m band. For a detailed discussion of SN\,2013df, see Szalai et al. (In Prep).

SN\,2013am and SN\,2011ja \citep{andrews2015} re-brighten in different epochs. SN\,2013am re-brightens slightly in the 4.5 $\mu$m band at around 370 days post-discovery, but not in the 3.6 $\mu$m band. SN\,2011ja stays luminous from 100 to 400d with higher fluxes at 400d. The data seem to suggest a re-brightening epoch in this time span. The secondary peak in the mid-IR has been observed in other Type II events, but at later times. \cite{meikle2011} and \cite{szalai2011} observed a re-brightening of SN\,2004dj (II-P) in all IRAC bands, except the 4.5 $\mu$m, in the 16 $\mu$m and 22 $\mu$m bands of IRS Peak-Up Imaging (IRS PUI) and 24 $\mu$m MIPS at 450-800 days post-explosion with the shorter wavelengths peaking first. The absence of a secondary peak in the 4.5 $\mu$m band was attributed to a strong contribution from CO 1-0 vibrational band at 4.65 $\mu$m before 500d. \cite{fabbri2011} reported a secondary mid-IR peak for SN\,2004et (II-P) in all IRAC bands, 16$\mu$m IRS-PUI, and 24 $\mu$m MIPS at 1000d. For both II-P events, attenuation in red-wing of spectral lines and an increase in optical light curve decline rates were observed as the mid-IR light curves started to rise again. This strongly suggested that the re-brightening was due to dust production in the ejecta of these SNe.
However, in the case of SN\,2013ai, where we do not have spectroscopic data, other sources of mid-IR flux such as additional heating mechanisms and CSM interactions cannot be ruled out. We also note that the amplitude of SN\,2013am's re-brightening is very small compared to that observed in SN\,2004dj and SN\,2004et. The interesting case of SN\,2011ja will be discussed in the next section. 

In addition to the re-brightening events, another outlier is evident in the light curves collage shown in Fig. \ref{fig:II_collage}. 
SN\,2014bi (II-P) is underluminous, especially in the 3.6 $\mu$m band in which it is 3-4 mags dimmer than other SNe. The color [3.6]-[4.5] is very high with a maximum at 2.95 mags at 264 days. This event will also be discussed in detail in Sec. \ref{2014bi}. 

\subsection{The Spectral Energy Distribution and Dust Parameter Fitting}

The Spectral Energy Distribution (SED) for each SN is fitted with an elementary one-component homogeneous dust model.  A more sophisticated model fitting would require additional data from other bands. The analysis follows the procedure described by \cite{fox2010, fox2011}. The flux density coming from warm dust at one equilibrium temperature is given by the Planck function modified with the dust mass absorption coefficient, which is frequency dependent. Assuming an optically thin dust shell with $\Delta r/r \sim 1/10$, the flux is given by 
\begin{equation} 
	F_\nu = \dfrac{M_{\rm d}B_{\nu } (T_{\rm d}) \kappa_\nu(a) }{d^2} 
	\label{dust_flux}
\end{equation}
where $B_{\nu } (T_{\rm d})$ is the Planck function, $\kappa_\nu(a) $ is the dust mass absorption coefficient as a function of the grain size $a$, $M_{\rm d}$ is the dust mass and $d$ is the distance of the source from the observer. The dust mass absorption coefficient is given by 
\begin{equation}\label{eq:kappa}
	\kappa_\nu(a) = \left(\dfrac{3}{4\pi\rho a^3}\right) \left(\pi a^2 Q_\nu(a)\right)
\end{equation}
where $\rho$ is the volume density of the dust and $Q_\nu(a)$ is the frequency dependent emission efficiency of the dust. $Q_\nu(a)$ is given in Fig. 4 of \cite{fox2010} and it is computed numerically. Because we only have two data points to fit to for each SN at each epoch, we are limited to using one temperature component and one dust composition. \cite{fox2011} found that assuming a pure graphite dust population with $a = 0.1 \mu$m gives the best fitting coefficients. The distinction between a graphite and a silicate dust population is a spectral feature around 9 $\mu$m, which is not captured by our data. They also found that a silicate dust model resulted in higher temperatures, masses, and luminosities. In many cases, silicate grains require an improbably high luminosity from the SN to heat them up to the observed temperature. Thus we assume a graphite dust population with a grain radius of 0.1 $\mu$m. We also note here that at early time ($\lesssim 200$d), the SN light still dominates dust emission in mid-IR and the results from SED fitting in those epochs might not reflect dust properties accurately.

\begin{figure*}
	\includegraphics[height = 7.5in]{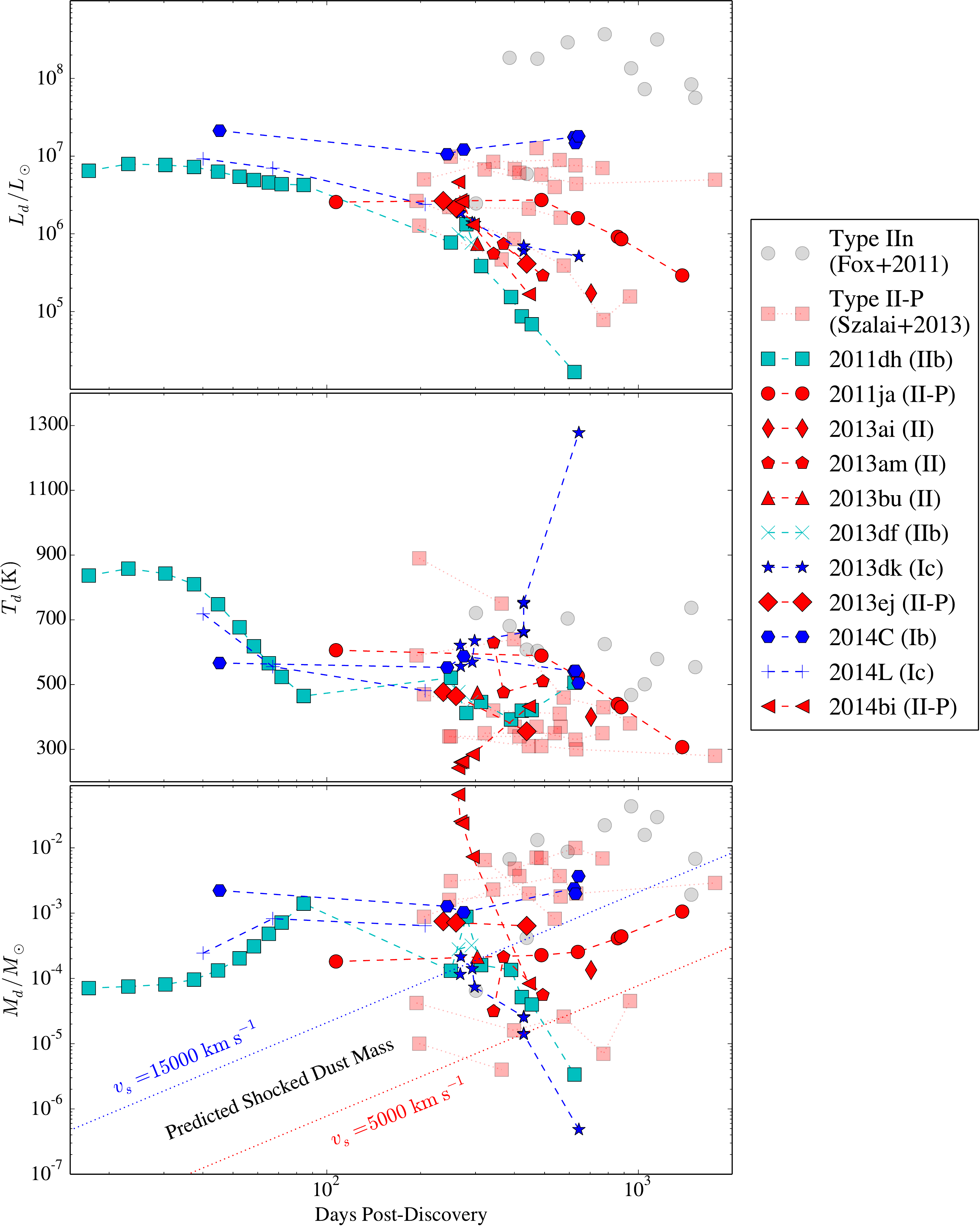}
	\caption{Time evolution of dust luminosity, temperature and mass from fitting equation \eqref{dust_flux} assuming 0.1 $\mu$m graphite grains. Shown on top of the dust mass evolution plot is the estimated dust mass assuming shock heating scenario using shock velocities of 5,000 $\rm km \ s^{-1}$ and 15,000 $\rm km \ s^{-1}$. The red points are Type II SNe, cyan points are Type IIb, and blue points are Type Ibc.The transparent grey circles are data for Type IIn SNe from \cite{fox2011}. The transparent red squares are data for Type II-P from \cite{szalai2013}. Data points with connecting dotted lines represent the same SN with multiple epochs of data. }
	\label{fig:ltd_evo}
\end{figure*}

We fit equation \eqref{dust_flux} to our data using the method \texttt{curve\_fit} in \texttt{SciPy} package, which varies fitting parameters to find the best fit to the data using the method of least squares \citep{jones}. Table \ref{dust_param} shows best fit temperatures and dust masses for all SNe for which data in both IRAC bands at the same epoch exist. The dust luminosity is computed by integrating equation \eqref{dust_flux} at all frequencies using the fitted dust mass and temperature. We note that due to the lack of spectral information, we cannot rule out preexisting dust and the best-fit dust masses reported here should be taken as an upper limit to the amount of newly formed dust. We find the dust masses that fall in the range $10^{-6}$ to $10^{-2}$ \msol\,, in agreement with previous work concluding that warm dust alone cannot account for all the dust predicted to be produced in a CCSN. Fig. \ref{fig:ltd_evo} shows the time evolution of dust luminosity, temperature and mass along with data on Type II-P SNe from \cite{szalai2013} and Type IIn SNe from \cite{fox2011}. The Type II-P events in \cite{szalai2013} and SNe in our sample have dust luminosity, temperatures, and mass estimates in the same range. In comparison with Type IIn events with confirmed early CSM interactions, SNe in our sample have less luminous dust emission, and at roughly the same temperature, less inferred dust mass. 

 In the plot showing dust mass (Fig. \ref{fig:ltd_evo}), two dotted lines represent the predicted dust mass that has been processed by the forward shock from the SN, which is given by Eq. 6 in \cite{fox2011}. Assuming the dust to gas ratio of 0.01, we get 
\begin{equation}
	M_{\rm d} (M_\odot) \approx 0.0028 \left(\dfrac{v_{\rm s}}{15,000 \ {\rm km \ s^{-1}}}\right)^3 \left(\dfrac{t}{\rm yr}\right)^2 \left(\dfrac{a}{\mu \rm m}\right)
\end{equation} where $v_{\rm s}$ is the shock velocity, $t$ is the time post-explosion, and $a$ is the grain size, assumed in this work to be $0.1 \ \mu$m. Comparing this prediction to the data, shock heating of pre-existing dust is likely ruled out for all events within a year following the explosion. Even for later times, to invoke shock heating we must assume a high shock velocity of 15,000 km $\rm s^{-1}$, which is unlikely for these events. 

The temporal evolution of the observed mass seems to show two different trends at  about a year post-explosion. Some of the events such as SN\,2011dh, SN\,2013dk, and especially SN\,2014bi, show a large decrease in the dust mass at this epoch. This could be a signature of either dust destruction or a reduced CSM/shock interactions.
Another group of SNe, especially SN\,2011ja and SN\,2014C show an increase in the dust mass, which could be due to either the formation of new dust or an intensifying dust heating source. 
For SN\,2011ja, both the dust luminosity and dust temperature are dropping arguing against the fading dust heating source scenario. 

\newpage
\begin{longtable*}{llllllllllll}
	\caption{A table of all epochs of observations with two simultaneous channels with image subtraction photometry along with the dust parameters from SED fitting.\protect\footnotemark{}}
	\label{dust_param} \\	
	\toprule
	MJD  & Epochs & $F_{\rm [3.6]}$ & Error  & $F_{\rm [4.5]}$  &  Error &$r_{\rm s}$ &$r_{\rm bb}$ & $L_{\rm d}$ & $M_{\rm d}$  & $T_{\rm d}$ \\
	& days& mJy & & mJy & & cm& cm& \lsol&\msol & K & \\ \midrule
 \multicolumn{11}{c}{SN\,2011dh (IIb)} \\ \midrule 
55730  &  17  &  5.309  &  0.004  &  4.87  &   0.01  &  $ 7.44 \times 10^{14}$  &  $ 8.45 \times 10^{15}$  &   $ 6.47 \times 10^{6}$  &  $ 7.10 \times 10^{-5}$  &  836 \\ 
55736  &  23  &  6.344  &  0.006  &  5.69  &   0.01  &  $ 9.96 \times 10^{14}$  &  $ 8.88 \times 10^{15}$  &   $ 7.93 \times 10^{6}$  &  $ 7.51 \times 10^{-5}$  &  858 \\ 
55743  &  30  &  6.261  &  0.006  &  5.71  &   0.01  &  $ 1.31 \times 10^{15}$  &  $ 9.07 \times 10^{15}$  &   $ 7.69 \times 10^{6}$  &  $ 8.08 \times 10^{-5}$  &  842 \\ 
55750  &  37  &  6.104  &  0.006  &  5.78  &   0.02  &  $ 1.62 \times 10^{15}$  &  $ 9.54 \times 10^{15}$  &   $ 7.25 \times 10^{6}$  &  $ 9.60 \times 10^{-5}$  &  809 \\ 
55758  &  44  &  5.556  &  0.006  &  5.70  &   0.01  &  $ 1.93 \times 10^{15}$  &  $ 1.04 \times 10^{16}$  &   $ 6.32 \times 10^{6}$  &  $ 1.32 \times 10^{-4}$  &  747 \\ 
55765  &  52  &  4.808  &  0.005  &  5.51  &   0.01  &  $ 2.27 \times 10^{15}$  &  $ 1.18 \times 10^{16}$  &   $ 5.43 \times 10^{6}$  &  $ 2.03 \times 10^{-4}$  &  676 \\ 
55771  &  58  &  4.16  &  0.01  &  5.33  &   0.01  &  $ 2.52 \times 10^{15}$  &  $ 1.35 \times 10^{16}$  &   $ 4.92 \times 10^{6}$  &  $ 3.10 \times 10^{-4}$  &  618 \\ 
55778  &  65  &  3.52  &  0.01  &  5.09  &   0.01  &  $ 2.81 \times 10^{15}$  &  $ 1.55 \times 10^{16}$  &   $ 4.58 \times 10^{6}$  &  $ 4.82 \times 10^{-4}$  &  565 \\ 
55785  &  71  &  2.94  &  0.01  &  4.77  &   0.01  &  $ 3.09 \times 10^{15}$  &  $ 1.77 \times 10^{16}$  &   $ 4.35 \times 10^{6}$  &  $ 7.18 \times 10^{-4}$  &  523 \\ 
55797  &  84  &  2.145  &  0.008  &  4.23  &   0.01  &  $ 3.64 \times 10^{15}$  &  $ 2.22 \times 10^{16}$  &   $ 4.25 \times 10^{6}$  &  $ 1.40 \times 10^{-3}$  &  464 \\ 
55963  &  250  &  0.517  &  0.003  &  0.845  &   0.004  &  $ 1.08 \times 10^{16}$  &  $ 7.52 \times 10^{15}$  &   $ 7.73 \times 10^{5}$  &  $ 1.31 \times 10^{-4}$  &  521 \\ 
55993  &  280  &  0.441  &  0.004  &  1.087  &   0.005  &  $ 1.21 \times 10^{16}$  &  $ 1.58 \times 10^{16}$  &   $ 1.33 \times 10^{6}$  &  $ 8.76 \times 10^{-4}$  &  412 \\ 
56026  &  312  &  0.171  &  0.002  &  0.363  &   0.002  &  $ 1.35 \times 10^{16}$  &  $ 7.24 \times 10^{15}$  &   $ 3.85 \times 10^{5}$  &  $ 1.60 \times 10^{-4}$  &  446 \\ 
56103  &  390  &  0.0413  &  0.0008  &  0.112  &   0.001  &  $ 1.69 \times 10^{16}$  &  $ 5.92 \times 10^{15}$  &   $ 1.54 \times 10^{5}$  &  $ 1.35 \times 10^{-4}$  &  392 \\ 
56135  &  422  &  0.0309  &  0.0007  &  0.074  &   0.001  &  $ 1.82 \times 10^{16}$  &  $ 3.89 \times 10^{15}$  &   $ 8.68 \times 10^{4}$  &  $ 5.17 \times 10^{-5}$  &  419 \\ 
56168  &  454  &  0.0248  &  0.0006  &  0.0586  &   0.0009  &  $ 1.96 \times 10^{16}$  &  $ 3.43 \times 10^{15}$  &   $ 6.86 \times 10^{4}$  &  $ 3.98 \times 10^{-5}$  &  421 \\ 
56337  &  623  &  0.0105  &  0.0004  &  0.0179  &   0.0005  &  $ 2.69 \times 10^{16}$  &  $ 1.17 \times 10^{15}$  &   $ 1.67 \times 10^{4}$  &  $ 3.36 \times 10^{-6}$  &  505 \\ 
\midrule \multicolumn{11}{c}{SN\,2011ja (II-P)} \\ \midrule 
56012  &  106  &  5.88  &  0.02  &  7.73  &   0.01  &  $ 4.62 \times 10^{15}$  &  $ 1.01 \times 10^{16}$  &   $ 2.57 \times 10^{6}$  &  $ 1.82 \times 10^{-4}$  &  606 \\ 
56393  &  488  &  6.07  &  0.01  &  8.29  &   0.02  &  $ 2.11 \times 10^{16}$  &  $ 1.10 \times 10^{16}$  &   $ 2.72 \times 10^{6}$  &  $ 2.27 \times 10^{-4}$  &  589 \\ 
56544  &  639  &  2.98  &  0.01  &  4.80  &   0.01  &  $ 2.76 \times 10^{16}$  &  $ 1.06 \times 10^{16}$  &   $ 1.59 \times 10^{6}$  &  $ 2.54 \times 10^{-4}$  &  526 \\ 
56764  &  859  &  1.07  &  0.02  &  2.365  &   0.009  &  $ 3.71 \times 10^{16}$  &  $ 1.19 \times 10^{16}$  &   $ 9.45 \times 10^{5}$  &  $ 4.51 \times 10^{-4}$  &  435 \\ 
56787  &  881  &  0.92  &  0.02  &  2.076  &   0.008  &  $ 3.81 \times 10^{16}$  &  $ 1.15 \times 10^{16}$  &   $ 8.46 \times 10^{5}$  &  $ 4.31 \times 10^{-4}$  &  430 \\ 
56912  &  1007  &  0.42  &  0.01  &  1.23  &   0.01  &  $ 4.35 \times 10^{16}$  &  $ 1.35 \times 10^{16}$  &   $ 6.81 \times 10^{5}$  &  $ 7.55 \times 10^{-4}$  &  376 \\ 
57288  &  1382  &  0.05  &  0.01  &  0.24  &   0.02  &  $ 5.97 \times 10^{16}$  &  $ 1.33 \times 10^{16}$  &   $ 2.91 \times 10^{5}$  &  $ 1.06 \times 10^{-3}$  &  306 \\ 
\midrule \multicolumn{11}{c}{SN\,2013ai (II)} \\ \midrule 
57057  &  704  &  0.0099  &  0.0002  &  0.0258  &   0.0003  &  $ 3.05 \times 10^{16}$  &  $ 6.02 \times 10^{15}$  &   $ 1.72 \times 10^{5}$  &  $ 1.34 \times 10^{-4}$  &  400 \\ 
\midrule \multicolumn{11}{c}{SN\,2013am (II)} \\ \midrule 
56715  &  342  &  0.0550  &  0.0004  &  0.0688  &   0.0005  &  $ 1.48 \times 10^{16}$  &  $ 4.36 \times 10^{15}$  &   $ 5.53 \times 10^{5}$  &  $ 3.14 \times 10^{-5}$  &  629 \\ 
56741  &  369  &  0.0461  &  0.0004  &  0.0873  &   0.0006  &  $ 1.59 \times 10^{16}$  &  $ 8.83 \times 10^{15}$  &   $ 7.38 \times 10^{5}$  &  $ 2.13 \times 10^{-4}$  &  475 \\ 
56865  &  493  &  0.0215  &  0.0003  &  0.0362  &   0.0004  &  $ 2.13 \times 10^{16}$  &  $ 4.80 \times 10^{15}$  &   $ 2.89 \times 10^{5}$  &  $ 5.56 \times 10^{-5}$  &  510 \\ 
\midrule \multicolumn{11}{c}{SN\,2013bu (II)} \\ \midrule 
56707  &  304  &  0.0925  &  0.0006  &  0.1755  &   0.0009  &  $ 1.31 \times 10^{16}$  &  $ 8.87 \times 10^{15}$  &   $ 7.44 \times 10^{5}$  &  $ 2.15 \times 10^{-4}$  &  475 \\ 
\midrule \multicolumn{11}{c}{SN\,2013df (IIb)} \\ \midrule 
56715  &  264  &  0.0512  &  0.0004  &  0.0963  &   0.0006  &  $ 1.14 \times 10^{16}$  &  $ 1.02 \times 10^{16}$  &   $ 9.93 \times 10^{5}$  &  $ 2.80 \times 10^{-4}$  &  477 \\ 
56741  &  291  &  0.0322  &  0.0003  &  0.0684  &   0.0005  &  $ 1.26 \times 10^{16}$  &  $ 1.02 \times 10^{16}$  &   $ 7.67 \times 10^{5}$  &  $ 3.19 \times 10^{-4}$  &  446 \\ 
56906  &  455  &  0.0088  &  0.0002  &  0.0147  &   0.0002  &  $ 1.97 \times 10^{16}$  &  $ 3.31 \times 10^{15}$  &   $ 1.43 \times 10^{5}$  &  $ 2.60 \times 10^{-5}$  &  515 \\ 
\midrule \multicolumn{11}{c}{SN\,2013dk (Ic)} \\ \midrule 
56733  &  268  &  0.1588  &  0.0007  &  0.2021  &   0.0009  &  $ 1.16 \times 10^{16}$  &  $ 8.26 \times 10^{15}$  &   $ 1.89 \times 10^{6}$  &  $ 1.16 \times 10^{-4}$  &  621 \\ 
56734  &  269  &  0.1378  &  0.0007  &  0.2040  &   0.0009  &  $ 1.16 \times 10^{16}$  &  $ 1.02 \times 10^{16}$  &   $ 1.85 \times 10^{6}$  &  $ 2.15 \times 10^{-4}$  &  556 \\ 
56758  &  292  &  0.1067  &  0.0006  &  0.1528  &   0.0008  &  $ 1.27 \times 10^{16}$  &  $ 8.44 \times 10^{15}$  &   $ 1.39 \times 10^{6}$  &  $ 1.41 \times 10^{-4}$  &  569 \\ 
56763  &  298  &  0.1164  &  0.0006  &  0.1440  &   0.0008  &  $ 1.29 \times 10^{16}$  &  $ 6.72 \times 10^{15}$  &   $ 1.36 \times 10^{6}$  &  $ 7.35 \times 10^{-5}$  &  635 \\ 
56894  &  428  &  0.0526  &  0.0004  &  0.0620  &   0.0005  &  $ 1.85 \times 10^{16}$  &  $ 4.13 \times 10^{15}$  &   $ 6.03 \times 10^{5}$  &  $ 2.59 \times 10^{-5}$  &  660 \\ 
57109  &  644  &  0.0198  &  0.0003  &  0.0133  &   0.0002  &  $ 2.78 \times 10^{16}$  &  $ 1.02 \times 10^{15}$  &   $ 5.11 \times 10^{5}$  &  $ 4.84 \times 10^{-7}$  &  1277 \\ 
\midrule \multicolumn{11}{c}{SN\,2013ej (II-P)} \\ \midrule 
56734  &  236  &  0.778  &  0.002  &  1.465  &   0.002  &  $ 1.02 \times 10^{16}$  &  $ 1.66 \times 10^{16}$  &   $ 2.65 \times 10^{6}$  &  $ 7.50 \times 10^{-4}$  &  477 \\ 
56758  &  259  &  0.587  &  0.001  &  1.158  &   0.002  &  $ 1.12 \times 10^{16}$  &  $ 1.58 \times 10^{16}$  &   $ 2.16 \times 10^{6}$  &  $ 7.15 \times 10^{-4}$  &  464 \\ 
56936  &  438  &  0.0363  &  0.0004  &  0.1224  &   0.0007  &  $ 1.89 \times 10^{16}$  &  $ 1.18 \times 10^{16}$  &   $ 4.13 \times 10^{5}$  &  $ 6.39 \times 10^{-4}$  &  355 \\ 
\midrule \multicolumn{11}{c}{SN\,2014C (Ib)} \\ \midrule 
56707  &  45  &  4.560  &  0.004  &  6.236  &   0.005  &  $ 1.96 \times 10^{15}$  &  $ 3.31 \times 10^{16}$  &   $ 2.44 \times 10^{7}$  &  $ 2.04 \times 10^{-3}$  &  588 \\ 
56905  &  243  &  2.474  &  0.003  &  3.134  &   0.004  &  $ 1.05 \times 10^{16}$  &  $ 2.12 \times 10^{16}$  &   $ 1.26 \times 10^{7}$  &  $ 7.52 \times 10^{-4}$  &  623 \\ 
56937  &  274  &  2.878  &  0.003  &  3.527  &   0.004  &  $ 1.19 \times 10^{16}$  &  $ 2.15 \times 10^{16}$  &   $ 1.44 \times 10^{7}$  &  $ 7.40 \times 10^{-4}$  &  640 \\ 
57284  &  622  &  3.551  &  0.004  &  5.076  &   0.005  &  $ 2.69 \times 10^{16}$  &  $ 3.18 \times 10^{16}$  &   $ 1.98 \times 10^{7}$  &  $ 1.99 \times 10^{-3}$  &  570 \\ 
57290  &  628  &  3.284  &  0.003  &  4.476  &   0.004  &  $ 2.71 \times 10^{16}$  &  $ 2.79 \times 10^{16}$  &   $ 1.75 \times 10^{7}$  &  $ 1.45 \times 10^{-3}$  &  590 \\ 
57304  &  642  &  3.340  &  0.003  &  5.227  &   0.005  &  $ 2.77 \times 10^{16}$  &  $ 3.66 \times 10^{16}$  &   $ 2.05 \times 10^{7}$  &  $ 2.95 \times 10^{-3}$  &  535 \\ 
\midrule \multicolumn{11}{c}{SN\,2014L (Ic)} \\ \midrule 
56723  &  40  &  1.152  &  0.002  &  1.234  &   0.002  &  $ 1.73 \times 10^{15}$  &  $ 1.37 \times 10^{16}$  &   $ 9.23 \times 10^{6}$  &  $ 2.44 \times 10^{-4}$  &  718 \\ 
56750  &  67  &  0.738  &  0.002  &  1.095  &   0.002  &  $ 2.90 \times 10^{15}$  &  $ 1.99 \times 10^{16}$  &   $ 7.01 \times 10^{6}$  &  $ 8.23 \times 10^{-4}$  &  555 \\ 
56890  &  206  &  0.1874  &  0.0008  &  0.348  &   0.001  &  $ 8.93 \times 10^{15}$  &  $ 1.55 \times 10^{16}$  &   $ 2.39 \times 10^{6}$  &  $ 6.43 \times 10^{-4}$  &  481 \\ 
\midrule \multicolumn{11}{c}{SN\,2014bi (II-P)} \\ \midrule 
56847  &  38  &  0.576  &  0.001  &  0.450  &   0.001  &  $ 1.67 \times 10^{15}$  &  $ 3.83 \times 10^{15}$  &   $ 2.90 \times 10^{6}$  &  $ 1.04 \times 10^{-5}$  &  1015 \\ 
57072  &  264  &  0.0106  &  0.0002  &  0.1060  &   0.0007  &  $ 1.14 \times 10^{16}$  &  $ 8.89 \times 10^{16}$  &   $ 4.94 \times 10^{6}$  &  $ 7.24 \times 10^{-2}$  &  241 \\ 
57076  &  268  &  0.0129  &  0.0002  &  0.1003  &   0.0007  &  $ 1.16 \times 10^{16}$  &  $ 5.64 \times 10^{16}$  &   $ 2.70 \times 10^{6}$  &  $ 2.54 \times 10^{-2}$  &  260 \\ 
57080  &  272  &  0.0130  &  0.0002  &  0.0996  &   0.0006  &  $ 1.18 \times 10^{16}$  &  $ 5.45 \times 10^{16}$  &   $ 2.59 \times 10^{6}$  &  $ 2.36 \times 10^{-2}$  &  261 \\ 
57102  &  293  &  0.0141  &  0.0002  &  0.0840  &   0.0006  &  $ 1.27 \times 10^{16}$  &  $ 3.27 \times 10^{16}$  &   $ 1.30 \times 10^{6}$  &  $ 7.31 \times 10^{-3}$  &  284 \\ 
57253  &  445  &  0.0210  &  0.0003  &  0.0470  &   0.0004  &  $ 1.92 \times 10^{16}$  &  $ 5.07 \times 10^{15}$  &   $ 1.67 \times 10^{5}$  &  $ 8.32 \times 10^{-5}$  &  432 \\ 
\midrule
\end{longtable*}
\addtocounter{footnote}{-2}	
\stepcounter{footnote}\footnotetext{All SN\,2011dh data is from PID 70207 (PI Helou) and the first 3 epochs of SN\,2011ja are from PID 80239 and 90178 (PI Andrews). The rest are from PID 10136 and 11063 (PI Kasliwal).}

\newpage
\section{Case Studies}\label{sec:cases}

\subsection{SN\,2011ja: A Type II-P with New Dust Formation on Cold Dense Shell}\label{2011ja}
SN\,2011ja is a Type II-P SN in the nearby galaxy NGC 4945 at a distance of 3.8 Mpc \citep{mould2008}. It was discovered on 2011 December 18 by \cite{monard2011} and the spectrum obtained a day later showed that it was a young Type II-P SN that shared similar spectral features with SN\,2004et a week after maximum light \citep{milisavljevic2011}. \cite{chakraborti2013} obtained X-ray and radio observations in two epochs out to about 100d. They found signatures of circumstellar interaction which likely resulted from the fast moving ejecta colliding with the slower stellar wind from the progenitor before it exploded. They deduced from the observations that the progenitor was likely a red supergiant with initial mass $M \gtrsim 12$ \msol. \cite{andrews2015} performed multi-band follow-up observations in the optical, near-IR, and mid-IR. The first 4 out of 10 epochs of the \textit{Spitzer} observations used in this work are also presented in their paper. After 105d, they observed a decrease in the optical flux, a rise in near-IR flux, and attenuation in the red-wing of H$\alpha$ emission lines. These are strong indications that SN\,2011ja is producing dust early on in the cold dense shell (CDS) created when the ejecta run into the CSM. Using a radiative transfer model, they inferred $\sim 10^{-4}$ \msol\, of dust at 100d, which is likely pre-existing dust heated by the SN flash. In the later epochs, they estimated that $\sim 10^{-5}$ \msol\ of dust being formed in the CDS. The mass of newly formed dust is an order of magnitude smaller than the pre-existing dust observed at earlier epochs. They classified SN\,2011ja with a growing group of Type II-P SNe, among them SN\,2004dj and SN\,2007od, which exhibit unusual evolution at late times due to the CSM interactions. This indicates that their progenitors lose enough mass before the SN explosion to create a dense CSM environment that is close enough for the forward shock to encounter within a few hundred days but not so that CSM interactions commence immediately after explosion as in the case of Type IIn SNe.

SN\,2011ja has been observed in 10 epochs by \textit{Spitzer}, 6 of which have concurrent coverage in both bands. The most notable features of the light curves are the almost constant luminosity out to 500d post explosion and the increasing [3.6]-[4.5] color. This evolution is consistent with a dust mass that increases monotonically with time with the declining temperature and dust luminosity.  
The dust mass increases from $10^{-4}$\msol\, in the first epoch at 107d to $10^{-3}$\msol\, in the last epoch at 1383d. 
Consider the blackbody radius which serves as a lower limit for the size of the dust shell responsible for the thermal emission in comparison with the shock radius. For every epoch of observations, we see that the shock radius, assuming the ejecta speed of 5,000 $\rm km \ s^{-1}$, is comparable to the blackbody radius. Both radii are within $10^{15}-10^{16}$ cm. This is consistent with the scenario proposed by \cite{andrews2015} in which dust grains condense in the cold dense shell (CDS) between the forward and reverse shocks. The inner and outer circumstellar radii of this zone will be at $4.5 \times 10^{15}$ and $4.5\times 10^{16}$ cm as proposed by \cite{andrews2015}. 

\begin{figure}
	\includegraphics[width = \linewidth]{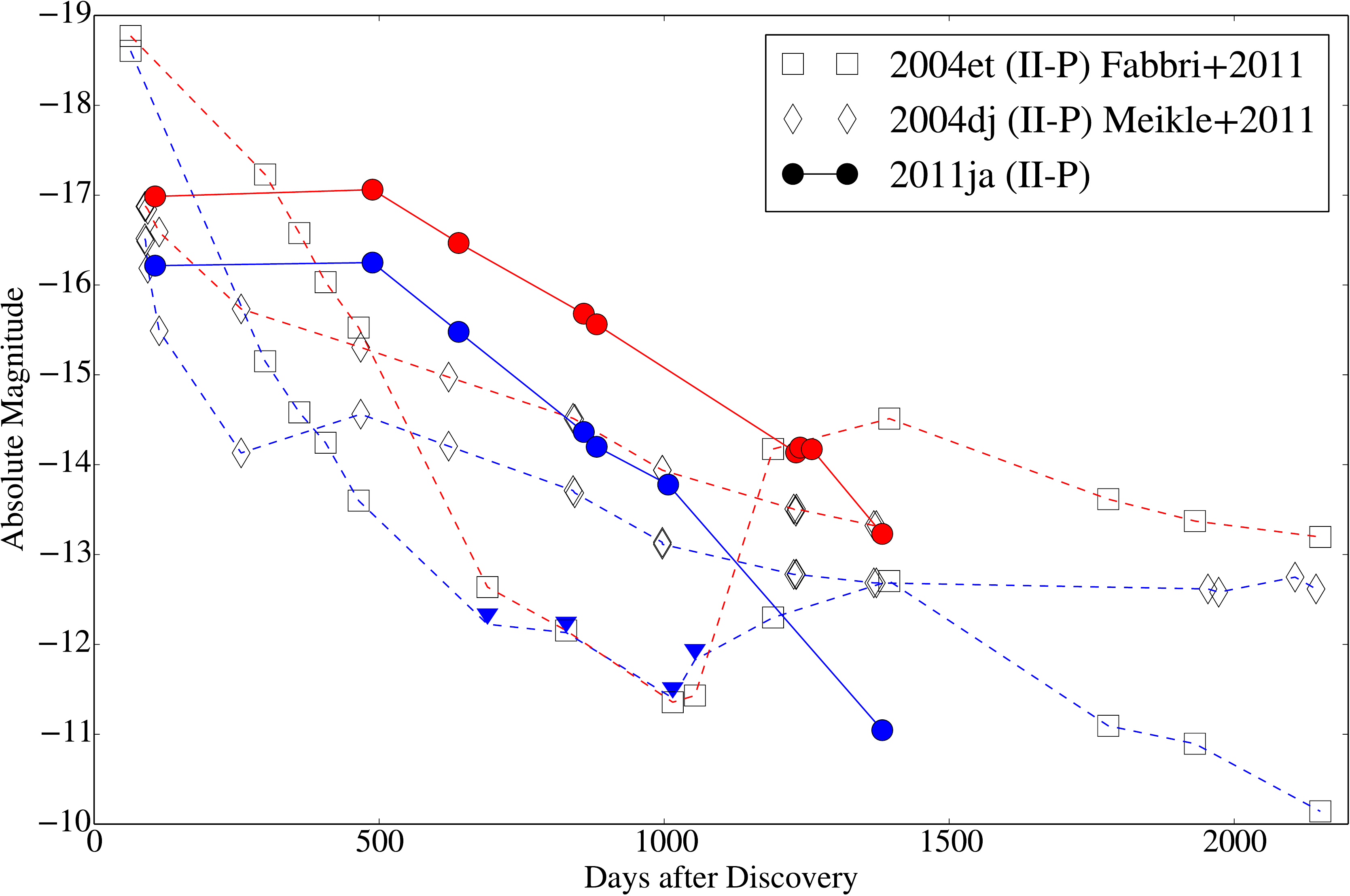}
	\caption{The light curve of SN\,2011ja in both IRAC channels, in comparison with that of SN\,2004et \citep{fabbri2011} and SN\,2004dj \citep{meikle2011,szalai2011}. SN\,2004et was not detected in the 3.6 $\mu$m band from around 700d to 1100d. The four downward arrows in the light curve of SN\,2004et indicate the three sigma detection limits. } 
	\label{fig:2011ja_vs_lits}
\end{figure}

In comparison with other Type II events in our sample, SN\,2011ja's early luminosity is very similar to that of SN\,2011dh (IIb) in both channels. However, SN\,2011ja's luminosity remains mostly constant out to 500d at which point it outshines every other SN in the sample. In comparison to the well-studied Type II-P SNe\,2004dj \citep{meikle2011, szalai2011} and SN\,2004et \citep{fabbri2011}, SN\,2011ja is more than an absolute magnitude brighter than both of them out to about 1,000d (see Fig. \ref{fig:2011ja_vs_lits}). 
In terms of dust properties, SN\,2004dj has $3.35 \times 10^{-5}$ \msol\, and $5.1 \times 10^{-5}$ \msol\, of dust at 859d and 996d respectively \citep{szalai2011}. 
For SN\,2004et, \cite{fabbri2011} used a more sophisticated clumpy dust model with an $R^{-2}$ density distribution, a range of grain sizes from 0.1 to 1 $\mu$m, and a graphite to silicate ratio of 1:4. Their largest mass estimate is $\sim 10^{-3}$ \msol\, at 690d. \cite{szalai2013}, using a graphite dust model, reported a maximum dust mass $10^{-2}$ \msol\, for SN\,2006bp at 628d. Most of the other Type II-P SNe in their sample produced around $5 \times 10^{-3}$ \msol\, of dust. The Type IIn SNe with strong CSM interactions tend to show more dust mass as shown in Fig. \ref{fig:ltd_evo}. \cite{fox2011, fox2013} reported the typical dust mass, assuming graphite dust composition, of at most $2.2 \times 10^{-2}$ \msol\ with most of their sample having an observed dust mass of around $5 \times 10^{-3}$ \msol\,. They conclude, as well, that most of the dust was pre-existing. 
A conclusion to draw from this comparison is that SN\,2011ja is not an outlier from the general population of dust producing Type II-P SNe. 


\subsection{SN\,2014C: An Interacting Stripped Envelope Supernova with a Possible LBV Projenitor}\label{2014C}
SN\,2014C is a Type Ib SN in the nearby galaxy NGC 7331, discovered on 2014 January 5. It was identified later on the same night as a young Type Ib SN with a spectrum matching that of SN\,1999ex at 6 days before maximum light \citep{kim2014}. The velocities of He and Si absorption features are measured to be about 13,800 km $\rm s^{-1}$ and 9,900 km $\rm s^{-1}$ respectively. SN\,2014C was observed with the Very Large Array (VLA) on 2014 January 11 to have a flux of $0.80 \pm 0.04$ mJy at 7 GHz \citep{kamble2014}. It was detected with by the Combined Array for Research in Millimeter Astronomy (CARMA) on 2014 January 13 at a flux level of $1.2\pm 0.2$ mJy in 85 GHz waveband \citep{zauderer2014}. \cite{milisavljevic2015} analyzed optical spectra of SN\,2014C and reported a remarkable metamorphosis of this SN from a typical Type Ib to an interacting Type IIn.\footnote{SN\,2001em underwent the similar transition from the Type Ibc to Type IIn when H$\alpha$, X-ray, and radio emissions were detected $\sim$ 2.5 years post-explosion \citep{chugai2006}. Nonetheless, for SN\,2014C, it was the first time that this transition was caught in action.}  
The onset of H$\alpha$ emission with intermediate line width along with an X-ray detection by \textit{Chandra} and the aforementioned radio emission are strong evidence of CSM interactions with a dense H-rich shell. The observations placed the shell radius at $\gtrsim 3 \times 10^{16}$ cm. The density of the unshocked CSM is constrained by the relative strength of the narrow lines in the spectra. The absence of a [O II] line in the presence of [O III] lines set a lower limit on the density of $10^4 \ \rm cm^{-3}$. The relative strengths of the [Fe VII] lines sets the upper limit of the density at $10^7 \ \rm cm^{-3}$. Furthermore, the width of the unresolved [O III] line indicates that the CSM velocity is less than 100 km $\rm s^{-1}$. Interestingly, no blueshifted line profile were reported, indicating that new dust did not form behind the shock. The earliest signature of CSM interactions observed from the ground came 113d post-explosion when the SN emerged from behind the Sun. The only observation carried out before this gap was on 2014 January 9, four days before maximum in V band. The gap in ground-based observations is filled in with \textit{Spitzer} observations.  

\begin{figure}
\includegraphics[width = \linewidth]{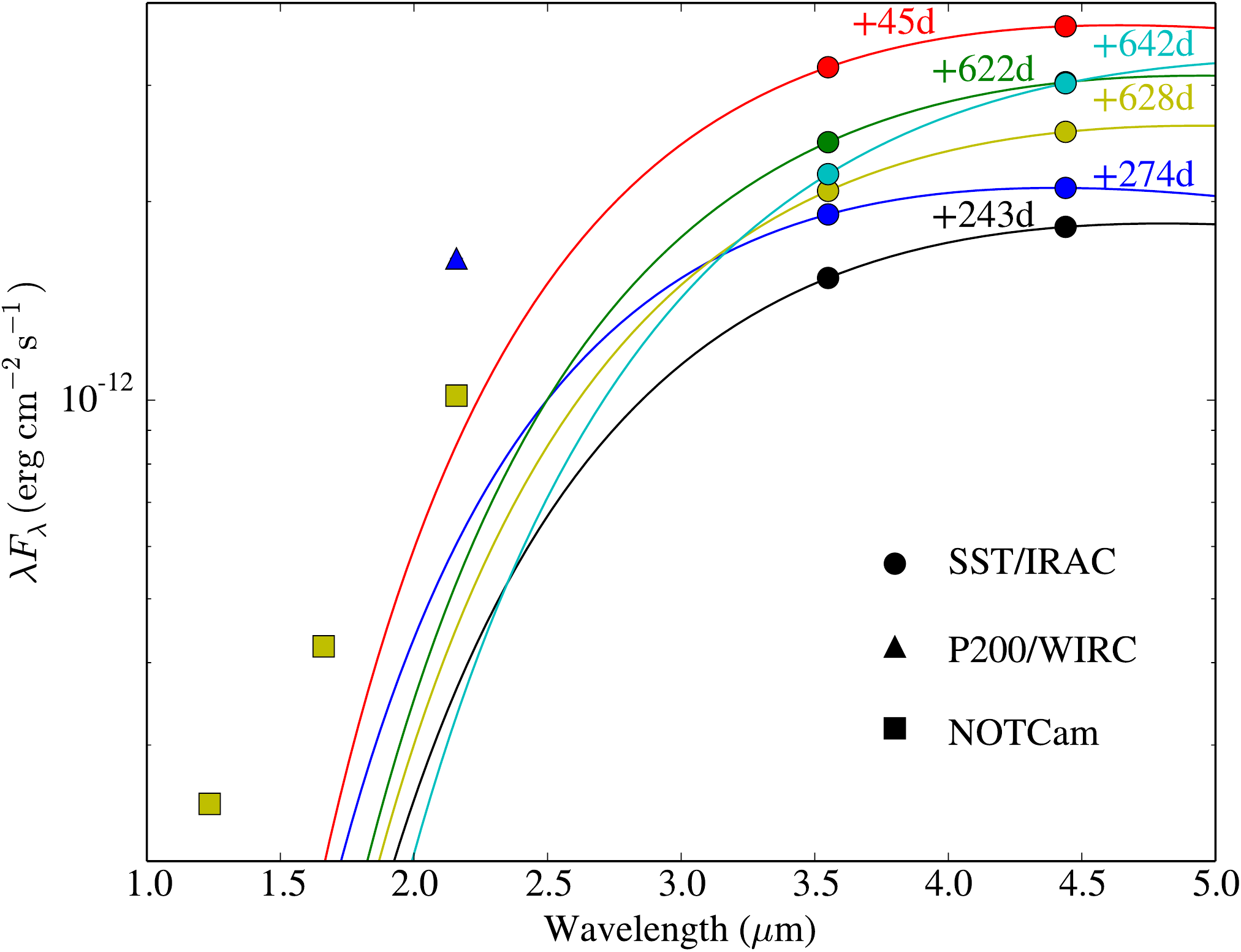}
\caption{An SED of SN\,2014C ranging from 45d to 642d post-explosion. \textit{Spitzer}/IRAC data are plotted in circles. Near-IR follow up observations obtained at the Nordic Optical Telescope near-infrared Camera and spectrograph (NOTCam) are plotted in squares, and the data obtained using the Wide-field InfraRed Camera (WIRC) on the 200-inch Telescope at Palomar Observatory are plotted in triangles. The overplotted lines are the best fit of equation \eqref{dust_flux} using only mid-IR data.}
\label{fig:2014C_nir_sed}
\end{figure}

SN\,2014C was observed in 6 epochs by \textit{Spitzer}, with the first epoch at 45d while the SN was behind the Sun for ground-based observers. 
The light curves in both channels were rising from $\sim$250 days until the latest epoch at $\sim$600d. Fig. \ref{fig:2014C_nir_sed} shows the time evolution of the SED of SN\,2014C along with concurrent NIR follow-up photometry at 274d and 628d post-explosion. The NIR fluxes indicate that the hot component of the SN has already faded, especially at 628d, and the SED is dominant by warm dust emission in mid-IR. Fluxes in both IRAC bands rise between 243d and 622d.  
While this re-brightening in mid-IR has been observed in a number of other events, even in some Type Ia with CSM interactions (see SN\,2005gj and SN\,2002ic; \citealp{fox2013}), it has never been observed directly in a stripped envelope CCSN. 
Multiple mechanisms could cause an SN to re-brighten in mid-IR at late times. In Type II SNe such as SN\,2004dj and SN\,2004et discussed earlier, the re-brightening can also be attributed to new dust being formed in either the SN ejecta or the CDS behind the shock. However, due to the lack of dust attenuation in optical spectral lines, this scenario is unlikely. In the cases of Type Ia-CSM events where dust production is not expected, \cite{fox2013} showed that the re-brightening can be attributed to pre-existing dust shell being radiatively heated by optical and X-Ray photons from CSM interactions. Another possible scenario is the collisional heating of pre-existing dust by the forward shock. We consider these last two shock heating mechanisms. 

Assuming a gas to dust mass ratio of 100 and that the dust sputtering timescales are significantly shorter than the cooling timescale of the shocked CSM, Fig. \ref{fig:ltd_evo} shows that the observed dust mass for SN\,2014C is always higher than the values predicted by shock heating. 
At +45 d, the blackbody radius, which gives the lower limit for the dust shell size, is a factor of 17 larger than the shock radius assuming shock velocity of $v_s = 5,000$ $\mathrm{km}\, \mathrm{s}^{-1}$. Even if we assume a maximum shock velocity of $v_s = 15,000$ $\mathrm{km}\, \mathrm{s}^{-1}$, the blackbody radius is still a factor of 5 larger. This is an indication that, at least for this earliest epoch, the observed mid-IR flux is likely due to pre-existing dust radiatively heated by CSM interactions. This requires that SN\,2014C starts to interact with the environment as early as 45 days post explosion. However, there are no X-Ray observations at that epoch to confirm the CSM interactions. \cite{milisavljevic2015} report an X-Ray luminosity of $1.3 \times 10^7$ $\mathrm{L}_\odot$ at 373 d, which can provide the dust luminosity observed around the same epochs.

We can estimate the CSM density making a simple assumption about the geometry for this scenario. At the 45 d epoch, the site of interaction is at the shock radius $r_s = 2.0 \times 10^{15}$ cm. The number density of hydrogen in the CSM can be approximated by

\begin{equation}
n_\mathrm{H}=\frac{3 M_\mathrm{d}}{4 \pi ((r_\mathrm{s}+r_\mathrm{d})^3-r_\mathrm{s}^3)}\frac{\chi}{m_\mathrm{H}},
\label{eq:nH}
\end{equation}

where $r_d$ is the radial thickness of the mid-IR emitting dust and $\chi$ is the gas-to-dust mass ratio for which we adopt a value of 100. The thickness $r_d$ can be approximated assuming the dust in the CSM is in thermal equilibrium with the incident radiation from the shock and exhibits a temperature of 550 K, the mean inferred dust temperature over all epochs. We assume the CSM is optically thin and composed of uniform distribution of 0.1 $\mu$m-sized silicate-type grains. The approximate distance at which 0.1 $\mu$m silicates are in thermal equilibrium with optical and UV photons of luminosity $\sim10^7$ $\mathrm{L}_\odot$ (\citealp{milisavljevic2015}; this work) is $r_d\sim6\times10^{16}$ cm, consistent with the blackbody radius. From equation \eqref{eq:nH}, we therefore derive a gas number density of $n_\mathrm{H}\sim3\times10^5$ $\mathrm{cm}^{-3}$. This is consistent with the limits of $10^4-10^7$ $\mathrm{cm}^{-3}$ given by \cite{milisavljevic2015}. Furthermore, taking the velocity of the CSM of $100$ $\mathrm{km}\, \mathrm{s}^{-1}$ to be constant, and the blackbody radius as an approximate size of the CSM shell, this material must have been ejected from the progenitor of SN\,2014C at least 100 years before the explosion. 

\cite{milisavljevic2015} propose three possible origins of the CSM shell around SN\,2014C: a short-lived Wolf-Rayet (WR) phase that swept up a slower red supergiant (RSG) wind, a single eruptive ejection from a luminous blue variable (LBV) or binary interaction, or externally influenced CSM confinement. All of these scenarios require a brief ($10^3$ yr) phase after the hydrogen envelope has been stripped. Although a thorough analysis of each scenario is beyond the scope of this paper, we present brief discussions on the implications of our results on the binary interaction and LBV eruption scenarios.

A possible origin for the dense dusty CSM of SN\,2014C is an outflow from a massive binary formed as the result of non-conservative mass transfer during a brief Roche-lobe overflow phase \citep{vanbeveren1998}. Such an event is thought to produce stripped-envelope WRs and is claimed to be occurring in the interacting binary system NaSt1 (e.g. \citealp{mauerhan2015}). NaSt1 \citep{crowther1999} hosts a WR-like star surrounded by a dusty, hydrogen-rich shell and may therefore be an example of a possible SN\,2014C progenitor. The dust formation efficiency and mass loss properties from massive binaries during non-conservative mass transfer are, however, unclear due to the paucity of observations of these systems. 

We now consider the LBV eruption scenario. LBV nebulae exhibit outflow velocities of $\sim$100 km/s \citep{smith2014}, which are consistent with the observationally based constraints on the CSM velocity ($\lesssim$ 100 km/s; \citealp{milisavljevic2015}) and imply a dynamical timescale of $\lesssim$100 yr given the inferred CSM blackbody radius. This timescale agrees with the observed $\sim$ 1-10 yr eruptive phases of LBVs. Given the inferred dust mass of the CSM ($2\times 10^{-3}$ $\mathrm{M}_\odot$) and assuming a gas to dust mass ratio of 100, the implied mass loss rate during the formation of the CSM is therefore $\gtrsim2 \times 10^{-2}$  $\mathrm{M}_\odot$ $\rm yr^{-1}$ and consistent with observed and predicted LBV mass loss rate \citep{smith2006, kochanek2011}. Observations of dusty LBV nebulae typically reveal $\gtrsim$ 0.01 \msol\, \citep{smith2006} of dust; however, we note that several LBV nebulae such as G24.73+0.69, HD 168625, and Hen 3-519 exhibit dust masses similar to that of the CSM (0.003 – 0.007 \msol\,; \citealp{ohara2003}, \citealp{clark2003}, \citealp{smith1994}. The consistency between mass loss properties of an LBV outburst and the measured dust mass and radius of the CSM point to a single LBV-like eruption roughly 100 years pre-explosion as a plausible origin of the CSM shell around SN\,2014C.

\subsection{SN\,2014bi: The Extremely Low Color Temperature}\label{2014bi}
SN\,2014bi is a low-luminosity Type II-P SN detected in the galaxy NGC 4096, 11.3 Mpc away, on 2014 May 31 \citep{kumar2014}. At the first epoch at 39d post-explosion, the SED is still dominated by the SN light. SN\,2014bi is about 1.5 mags dimmer than SN\,2011dh at the same epoch in both channel with small [3.6]-[4.5] color. 200 days later, however, SN\,2014bi becomes a very red source with [3.6]-[4.5] $\sim$ 3. This color temperature is associated with a low blackbody temperature of around 200 K. 
The dust mass calculated for this event is pushing $6.57 \times 10^{-2}$ \msol\, at 264d, which is relatively high in comparison to dust masses for other CCSNe. However, the dust mass drops sharply to $2.54 \times 10^{-2}$ \msol\, only 4 days later. And by 293d, the dust mass drops further to $7.31 \times 10^{-3}$ \msol\,, an order of magnitude lower than that in the second epoch. At 445d, the dust is almost gone, with a residual value of only $8.32 \times 10^{-5}$ \msol\,. 

There are a number of possible causes for this drop in dust mass. Firstly, this could be a signature of the SN shock that overtakes and destroys the dust shell. If we assume this scenario we can calculate the density of the dust shell. To get an upper limit of the density, we assume a slow shock velocity of $v_s = 5,000 \ \rm km \  s^{-1}$. Assuming a grain size of 0.1 $\mu$m as before and and a grain density of 2$\rm \  g \  cm^{-3}$, the number density of dust particles needed to account for the observed decline in mass is only $3\times10^{-2} \  \rm cm^{-3}$ which is very low considering the typical density of a dust shell  $\sim 10^5 \ \rm cm^{-3}$. Another possible explanation is that the radiative heating source, e.g. emission from CSM interactions, is fading; thus the amount of heated dust decreases. 
Finally, due to the lack of data at shorter wavelengths, it is also possible that a hotter component of the SN has cooled down and started to emit more significantly in the 3.6 $\mu$m band. In this case, the SED fit will underestimate the dust mass. 
We also note that the 3.6 $\mu$m band can be contaminated by the PAH emission while the 4.5 $\mu$m band can be contaminated by CO emission at 4.65 $\mu$m. A complementary study is indeed needed in order to elucidate the temporal evolution of this SN.

\subsection{Supernovae with Long-lived Infrared Emission Decades After Explosion}\label{old}
\begin{figure}
	\includegraphics[width = \linewidth]{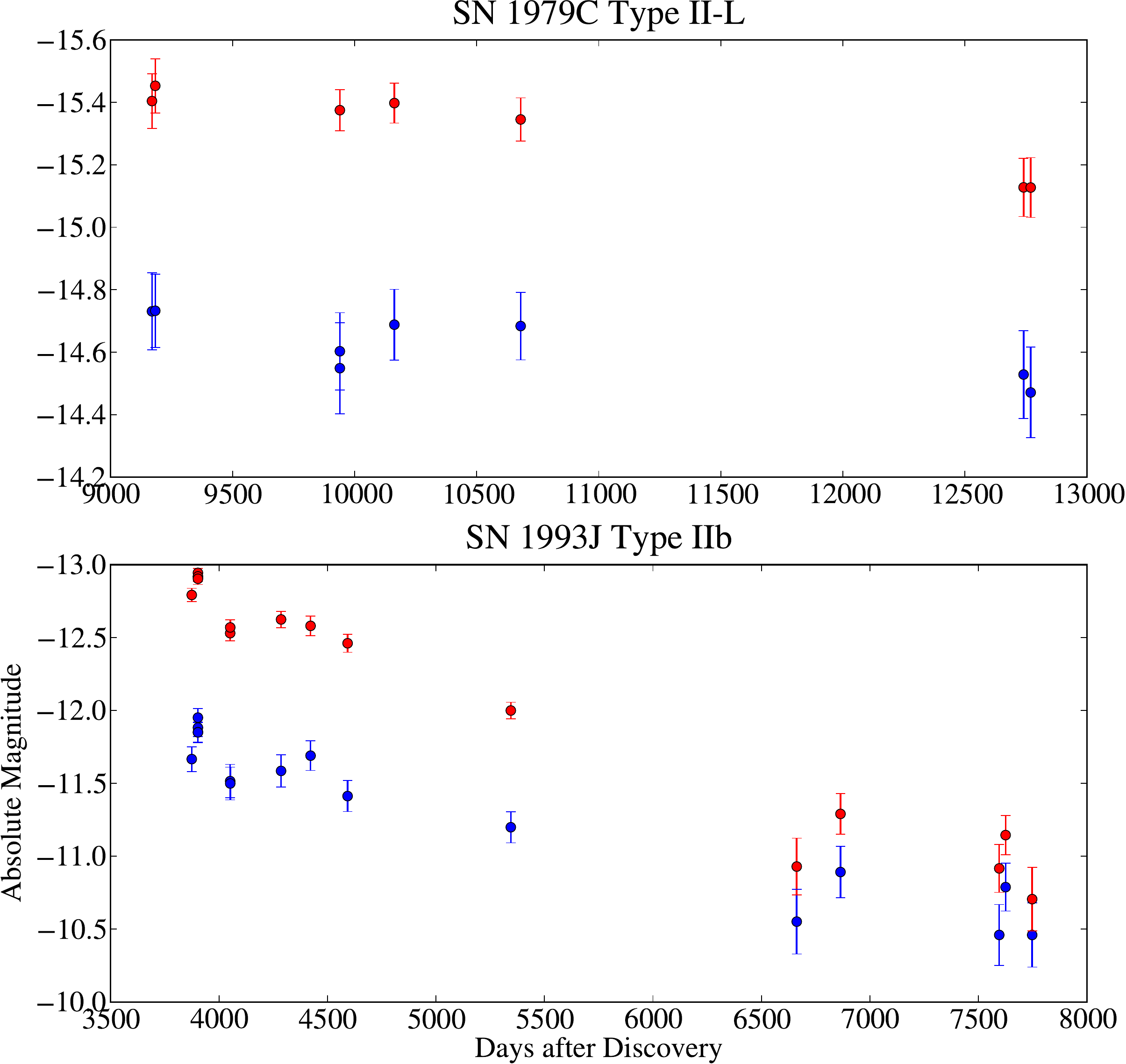}
	\caption{Two light curves show the very late-time evolution of SN\,1979C (top) and SN\,1993J (bottom). The data are aperture photometry from the science frames, so the background contribution has not been fully subtracted out. For SN\,1993J, the background emissions is minimal. For SN\,1979C, the photometry is affected by association with a cluster. The SN light alone indeed was not as bright as [4.5] $\sim$ -15 mag at such late epochs. }
	\label{fig:late_sne_lc}
\end{figure}

\begin{figure}
	\includegraphics[width = \linewidth]{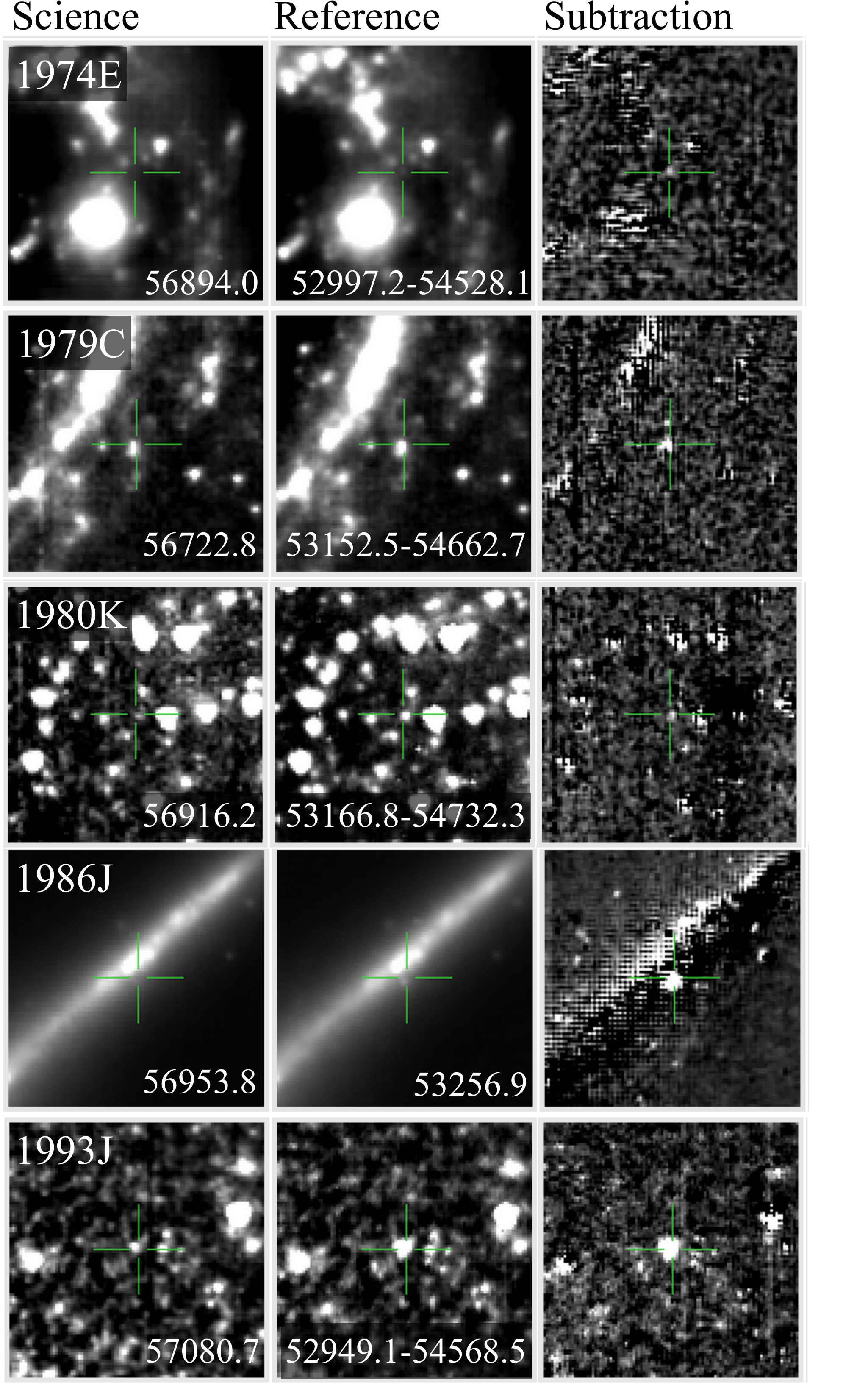}
	\caption{Image subtractions of five old Type II SNe showing the decrease in flux in the science frames taken at the later epochs compared to the references. From top to bottom, SN\,1974E (II), SN\,1979C (II-L), SN\,1980K (II-L), SN\,1986J (IIn), and SN\,1993J (IIb). The MJD of each observation is shown on top of the image.  In most cases, the references are stacks of \textit{Spitzer} observations over the MJD ranges shown.}
	\label{fig:old_typeII_figs}
	
\end{figure}

Surprisingly, we detect IR emissions from SN\,1974E (II), SN\,1979C (II-L), SN\,1980K (II-L), SN\,1986J (IIn), and SN\,1993J (IIb), more than 20 years after explosions. The extended light curves for SN\,1993J and SN\,1979C in Fig. \ref{fig:late_sne_lc} show that the luminosity for both events is still declining. This indicates that we are still detecting the light associated with the SN and not just a constant background source, the remnant, or the companions of the progenitors of those SNe. The notable features of these old SNe are that they have the color [3.6]-[4.5] $\sim$ 1 mag, and that the evolution is very slow. SN\,1993J faded by $\sim$1 mag over the course of a decade while the older SN\,1979C evolved even more slowly, fading only by $\sim$0.2 mag over a decade. For other events without enough epochs of data to construct good light curves, we present the image subtraction thumbnails which show that the SN has at least decayed from the reference epoch in Fig. \ref{fig:old_typeII_figs}. From the thumbnails, one can appreciate the importance of image subtraction. In most cases, it is not possible to determine from the science frames alone whether or not we have a detection. Even in cases where there is an obvious source at the location of the SN, it is still difficult to differentiate between the SN light and the surrounding contribution. Image subtraction using a baseline of almost a decade as used in this work uncovers a number of old SNe which have never been reported to have late-time mid-IR emission previously (see Table 1 in \cite{fox2011}). 

\section{Conclusion}
We have presented a study of the general population of nearby supernovae that fall within the coverage of the SPitzer InfraRed Intensive Transient Survey (SPIRITS). The main goal of this project is threefold: (1) to study the detection statistics for the whole SNe sample; (2) to generate light curves for Type Ib/c and Type II SNe in order to identify the norms and to uncover unusual events; and (3) to derive the dust parameters especially the dust mass associated with each event. The observations were made with the 3.6 $\mu$m and 4.5 $\mu$m channels of IRAC on board \textit{Spitzer}. The data, both from new SPIRITS observations in 2014-2015 and from other programs obtained through the \textit{Spitzer Heritage Archive}, are put through SPIRITS image subtraction pipeline. 

The detection statistics are obtained by visually vetting all observations of all SNe in the SPIRITS's footprint. The results show that Type Ia SNe fade away much more quickly than Type Ib/c and Type II core-collapse SNe, with none of the Type Ia living past 3 years. Type II SNe, on the other hand, tend to live for a very long time with 10 events living longer than 3 years. Five of those events, SN\,1974E (II), SN\,1979C (II-L) , SN\,1980K (II-L), SN\,1986J (IIn), and SN\,1993J (IIb), are detected decades after the explosion, providing the first sample of mid-IR observation of CCSNe at such late epochs. Other earlier surveys of SNe in the \textit{Spitzer} archive missed these events because image subtraction was not done to reveal the subtle time evolution on top of the bright background. 

We have created light curves collages for Type Ib/c and Type II SNe that have pre-explosion reference images using image subtraction based photometry. The light curves are heterogeneous compared to the light curves of Type Ia events. 
The SEDs of the events with image subtraction based photometry of both IRAC channels at the same epochs have been fitted fit with an elementary dust model with graphite grains of radius 0.1 $\mu$m at a single temperature. Dust luminosity, temperature, and mass are obtained from the fit. The light curves collages and temporal evolution of the dust parameters reveal some outliers from the general norm.
An outlier among the Type Ib/c SNe is the SN\,2014C, which rebrightens around 260d post explosion with increasing observed dust mass. The lack of signatures for dust formation in the optical spectra \citep{milisavljevic2015} suggests that the emission in this SN comes from pre-existing dust. The blackbody radius at the first epoch suggests that the dust shell is heated by emission from CSM interactions. From the dust mass and blackbody radius, we obtain an estimate for the gas density of $\sim 10^{6} \ \rm cm^{-3}$, which is consistent with an LBV origin.
The majority of the hydrogen-rich Type II SNe's light curves loosely follow that of SN\,2011dh, a well-observed Type IIb SN. We identify two interesting outliers. SN\,2011ja (II-P) is over-luminous at 1000 days post-discovery. The temporal evolution of its dust parameters seems to suggest either an episode of dust production near the shock front or the radiative heating of pre-existing dust by CSM interactions. The location of the emission at $\sim 10^{16}$ cm agrees with the location of new dust formation in the CDS proposed by \citep{andrews2015}. SN\,2014bi (II-P), on the other hand, is under-luminous and has a very low color temperature. The evolution of its dust parameters is opposite of that of other events, suggesting either dust destruction or dust heating with fading source. 
The dust mass determined for all events agree with the broad consensus that core-collapse SNe only produce $10^{-6}$ to $10^{-2}$ \msol\, of dust. 
This work demonstrates the power of mid-IR observations as a powerful diagnostic for examining unusual temporal evolution of CCSNe at late times. A multi-epoch survey such as SPIRITS targeting nearby CCSNe would be crucial in establishing the rate of these rare events.


This work made use of observations from the \textit{Spitzer Space Telescope} operated by the Jet Propulsion Laboratory, California Institute of Technology, under a contract with NASA. Ground-based observations presented were obtained at the Palomar Observatory, operated by California Institute of Technology. The Mount Lemmon Observing Facility is operated by the University of Minnesota. The Nordic Optical Telescope is operated by the Nordic Optical Telescope Scientific Association at the Observatorio del Roque de los Muchachos, La Palma, Spain, of the Instituto de Astrofisica de Canarias. We are grateful for support from the NASA Spitzer mission grants to the SPIRITS program (PIDs 10136 \& 11063). S.T. was supported by the Royal Thai Scholarship and a portion of this work was done while under the Claremont-Carnegie Astrophysics Research Program. R.D.G. was supported by NASA and the United States Air Force. 

\bibliographystyle{apj}
\bibliography{midir_sn_lightcurve_apj}
  
\end{document}